\documentclass[12pt]{article}
\usepackage{latexsym}
\usepackage{amsmath}
\usepackage{amssymb}
\usepackage{relsize}
\usepackage{geometry}
\usepackage{tensor}
\usepackage{physics} 
\usepackage{csquotes} 

\usepackage{graphicx}       	
\usepackage[font=small,labelfont=bf]{caption}
\usepackage{subcaption}

\def\beqn{\begin{eqnarray}}
\def\eeqn{\end{eqnarray}}

\def\beq{\begin{equation}}
\def\eeq{\end{equation}}
\def\ba{\beq\new\begin{array}{c}}
\def\ea{\end{array}\eeq}

\newcommand{\gsim}{\lower.7ex\hbox{$
\;\stackrel{\textstyle>}{\sim}\;$}}
\newcommand{\lsim}{\lower.7ex\hbox{$
\;\stackrel{\textstyle<}{\sim}\;$}}

\newcommand{\ntwo}{${\mathcal N}=2$ }

\newcommand{\ntwot}{${\mathcal N}= \left(2,2\right) $ }

\newcommand{\pt}{\partial}

\newcommand{\wcpt}{$\mathbb{WCP}(2,2)\;$}
\newcommand{\wcp}{$\mathbb{WCP}(N,\tilde N)\;$}
\newcommand{\cpone}{$\mathbb{CP}(1)\;$}

\numberwithin{equation}{section}

\newcommand{\p}{\partial}
\newcommand{\wt}{\widetilde}
\newcommand{\ov}{\overline}

\def\slashed#1{\setbox0=\hbox{$#1$}             
   \dimen0=\wd0                                 
   \setbox1=\hbox{/} \dimen1=\wd1               
   \ifdim\dimen0>\dimen1                        
      \rlap{\hbox to \dimen0{\hfil/\hfil}}      
      #1                                        
   \else                                        
      \rlap{\hbox to \dimen1{\hfil$#1$\hfil}}   
      /                                         
   \fi}                                        %

\newcommand{\WCP}{$\mathbb{WCP}(N,\tilde N)\;$}
\newcommand{\tN}{\widetilde{N}}

\usepackage{mathtools}
\usepackage{hyperref}
\usepackage{xcolor}
\usepackage{tikz-cd}
\usepackage{placeins}
\usepackage[toc,page, title, titletoc]{appendix}

\begin{document}

\hypersetup{%
	linkbordercolor=blue,
}

%
%

\begin{titlepage}

\begin{flushright}
FTPI-MINN-20-19, UMN-TH-3920/20\\
June 15, 2020
\end{flushright}


\begin{center}
	\Large{{\bf 
		String \boldmath{\textquote{Baryon}}   in Four-Dimensional \\ $\mathcal{N} = 2$ Supersymmetric QCD from 2D-4D Correspondence
		
	}}
	
\vspace{5mm}
	
{\large  \bf E.~Ievlev$^{\,a,b,c}$, M.~Shifman$^{\,d, e}$ and  A.~Yung$^{\,a,e}$}
\end{center}
\begin{center}

	$^{a}${\it National Research Center ``Kurchatov Institute'',
	Petersburg Nuclear Physics Institute, Gatchina, St. Petersburg
	188300, Russia}\\
	$^{b}${\it  St. Petersburg State University,
	 Universitetskaya nab., St.~Petersburg 199034, Russia}\\
$^{c}${\it Saint Petersburg State Electrotechnical University,
		ul. Professora Popova, St.~Petersburg 197376, Russia}\\
	{\it  $^{d}$Department of Physics,
University of Minnesota,
Minneapolis, MN 55455}\\{\small and}\\
{\it  $^{e}$William I. Fine Theoretical Physics Institute,
University of Minnesota,
Minneapolis, MN 55455}\\
	
	\end{center}

\vspace{5mm}

\begin{center}
{\large\bf Abstract}
\end{center}

We study non-Abelian vortex strings in four-dimensional (4D) $\mathcal{N} = 2$ supersymmetric QCD with U$(N=2)$ gauge group and $N_f=4$ flavors of quark hypermultiplets. It has been recently shown that these vortices behave as critical superstrings. The spectrum of closed string states  in the associated string theory was found and interpreted as a spectrum of hadrons in 4D $\mathcal{N} = 2$ supersymmetric QCD. In particular, the lowest  string state appears to be a massless BPS ``baryon." Here we show the occurrence  of this stringy baryon using a purely field-theoretic method. To this end we study the conformal world-sheet theory on the non-Abelian string -- the so called weighted $\mathcal{N} = (2,2)$ supersymmetric $\mathbb{CP}$ model. Its target space is given by the six-dimensional non-compact Calabi-Yau space $Y_6$, the conifold. We use  mirror description of the model to study  the 
BPS kink spectrum  and its transformations on curves (walls) of marginal stability. Then we use the 2D-4D correspondence to show that the deformation of the complex structure of the conifold  is associated with the emergence of a non-perturbative Higgs branch in 4D theory which opens up at strong coupling. The modulus parameter on this Higgs branch is 
the vacuum expectation value of the massless BPS ``baryon" previously found in string theory.

\end{titlepage}

\newpage

\tableofcontents

%
%

\section{Introduction \label{sec:introduction}}

In 2015 a non-Abelian semilocal vortex string was discovered possessing a world-sheet theory which is both superconformal and critical \cite{SYcstring}. 
This string is supported in four-dimensional (4D) ${\cal N} = 2$ super-QCD (SQCD) with the U$(N=2)$ gauge group, $N_f=4$ flavors 
of quarks and a Fayet-Iliopoulos (FI) term \cite{FI}.  
Due to the extended supersymmetry,  the gauge coupling  in the 4D bulk could be renormalized only at one loop.
With our judicial choice of the matter sector ($N_f=2N$) the one-loop renormalization cancels. No dynamical scale parameter 
$\Lambda$ is generated in the bulk \footnote{However, conformal invariance of 4D SQCD is broken by the Fayet-Iliopoulos term.}.

This is also the case in the world-sheet theory described by the weighted $\mathbb{CP}$  model ($\mathbb{WCP}(2,2)$), see Sec.~\ref{sec:wcp} below. Its $\beta$ function vanishes,  and the overall Virasoro central charge is critical \cite{SYcstring}.  
This happens because in
addition to four translational moduli, non-Abelian string has six orientational and size
moduli described by $\mathbb{WCP}(2,2)$ model. Together, they form a ten-dimensional target space required for a superstring to be critical. The target
space of the string sigma model is $\mathbb{R}_4\times Y_6$, a product of the flat four-dimensional space and a Calabi-Yau non-compact threefold $Y_6$, namely, the conifold.

This allows one to apply string theory for consideration of the closed string spectrum and its interpretation  as a spectrum of hadrons in 4D \ntwo SQCD. The vortex string at hand was identified as the string theory of Type IIA \cite{KSYconifold}.

The study of the above vortex string from the standpoint of string theory, with the focus on massless states in four dimensions has been started   in \cite{KSYconifold,KSYcstring}. Later the low lying massive string states were found by virtue of little string theory \cite{SYlittles}.
Generically,  most of massless modes have  non-normalizable wave functions over the conifold $Y_6$, i.e. they are not localized in 4D 
and, hence, cannot be interpreted as dynamical states in 4D SQCD.  In particular, no massless 4D gravitons or vector fields were found  in the physical spectrum in \cite{KSYconifold}. However, a single massless BPS hypermultiplet in the 4D bulk was detected 
at a self-dual point (at strong coupling). It is associated with deformations of a complex structure of the conifold and was  interpreted  as a composite 4D ``baryon.''\footnote{If the gauge group is U(2), as is our case, there are no {\em bona fide} baryons. We still use the term baryon because of a particular value of its  charge $Q_B$(baryon) = 2 with respect to the global unbroken U(1)$_B$, see Sec.~\ref{conifold}.} 

\begin{figure}[!t]
\centerline{\includegraphics[width=10cm]{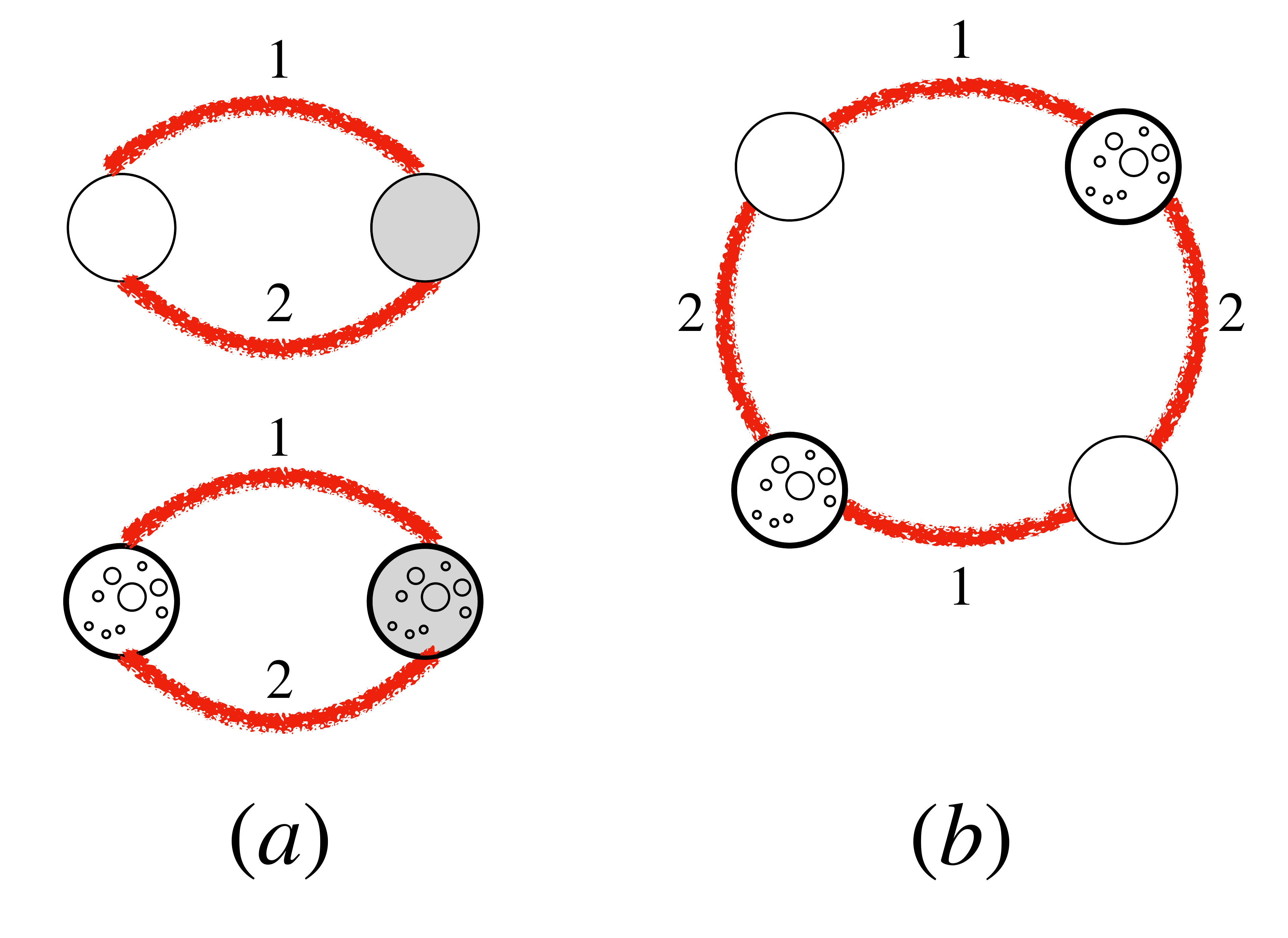}}
\caption{\small Examples of the monopole ``necklaces": (a) mesonic; (b) baryonic. 1,2 refer to two types of strings corresponding to two vacua on the 
string world sheet. The shaded circles are antimonopoles.
The two types of kinks are the $n^P$-kinks and $\rho^K$ kinks.}
\label{monmb} 
\end{figure}

Previous studies of the vortex strings  supported in four-dimensional  ${\cal N} = 2$ super-QCD at weak coupling showed that the non-Abelian vortices confine monopoles. The elementary monopoles are junctions of two distinct elementary non-Abelian strings \cite{SYmon,HT2}. In the 4D bulk theory 
we have monopole-antimonopole mesons in which monopole and antimonopole are connected by two confining strings 
(see Fig.~\ref{monmb}). For the U(2) gauge group we can have also ``baryons'' consisting of even number of  monopoles rather than of the monopole-antimonopole pair. 

The monopoles acquire quantum numbers with respect to the global group 
\begin{equation}
	{\rm SU(2)}\times{\rm SU(2)}\times{\rm U(1)}_B
	\label{globgroup_d=4}
\end{equation}
of the 4D SQCD, see \cite{SYrev} for a review. Indeed, in the world-sheet model on the vortex string, confined monopole are seen as kinks interpolating between two different vacua \cite{SYmon,HT2}. These kinks are described at strong coupling by  $n^P$ and $\rho^K$ fields \cite{W79,SYtorkink} (for \wcpt model $P=1,2$, $K=3,4$, see Sec.~\ref{sec:mirror}).
These two types of kinks correspond to two types of monopoles -- both have the same magnetic charge but different global charges. This is seen from the fact that
the global symmetry in the world-sheet theory on the string is exactly the same as given in Eq. (\ref{globgroup_d=4})
and the U(1) charges of the $n^P$ and $\rho^K$ fields are 0 and 1, respectively.
 One of them   is a fundamental field in the first SU(2) group and the other  in the second, 
\begin{equation}
	n^P \sim (\textbf{2},\,\textbf{1},\, 0), \qquad \rho^K \sim (\textbf{1},\,\textbf{2},\, 1)\,.
	\label{ksat}
\end{equation}
This refers to confined 4D monopoles too.

Our general strategy is as follows. We explore the BPS protected sector of the world-sheet model, two-dimensional \wcpt\!\!, starting from weak coupling $\beta \gg 1$, where $\beta$ is the inverse coupling. This procedure requires an infra-red (IR) regularization.  To this end we introduce  masses of quarks in 4D SQCD. They translates into four twisted masses in the world-sheet \wcpt (two for $n^P$  and two for $\rho^K$) which we arrange in a certain hierarchical order. We find both vacua of the theory, and study distinct kinks (in the mirror representation).  
Thus the  vacuum structure   and kink spectrum of this theory are  known exactly, and so are all curves (walls) of the marginal stability (CMS). Then we move towards strong coupling $\beta \sim 0$ 
carefully identifying CMS in the complex $\beta$ plane.
At each step we determine which kinks decay on CMS and which are stable upon crossing and establish their relation to four-dimensional monopoles using the so called 2D-4D correspondence, the coincidence of BPS spectra in 4D \ntwo SQCD and in the string world-sheet theory \cite{SYmon,HT2,Dorey}.

At strong coupling we use 2D-4D correspondence to confirm that our 4D SQCD enters the so called ``instead-of-confinement'' phase found earlier in  asymptotically free versions of SQCD \cite{SYdual}, see \cite{SYdualrev} for a review.
This phase  is qualitatively similar  to
the conventional QCD confinement: the quarks and gauge bosons screened at
weak coupling, at strong coupling evolve into monopole-antimonopole pairs
confined by non-Abelian strings. They form monopole mesons and baryons shown in Fig.~\ref{monmb}. The role of the constituent quark in this phase is played by the confined monopole.

Needless to say, the quark  masses break the global symmetry (\ref{globgroup_d=4}). At the very end we tend them to zero, restoring the global symmetries, as well as conformal invariance of the world-sheet theory on the string. 

Our main result is the emergence (at $\beta =0$) of a short BPS massless ``baryon'' supermultiplet with the U(1)$_{B}$ charge $Q_B=2$.  In this way we demonstrate that the massless ``baryon'' state which had been previously observed using string theory
arguments \cite{KSYconifold} is seen in the field-theoretical approach too. We believe this is the first example of this type.

To obtain this result we use the following strategy. It is known that \wcpt \\
model at $\beta=0$ has a marginal deformation
associated with the deformation of the complex structure of the conifold. Since \wcpt model is a world-sheet theory on the non-Abelian string the natural question to address is what is  the origin of this deformation in 4D SQCD. On general grounds one expects that this could be some parameter of the 4D theory such as a coupling constant. Another option is that it could be a modulus, a vacuum expectation value (VEV) of a certain dynamical field. We show using 2D-4D correspondence that the latter option is realized in the case at hand. A new non-perturbative Higgs branch opens up at $\beta=0$ in  4D SQCD. The modulus parameter on this Higgs branch is  the VEV of the massless BPS baryon
constructed from four monopoles connected by confining strings as shown in Fig.~\ref{monmb}b.

The organization of the paper is as follows.
Section \ref{SQCD} presents a brief review of four-dimensional {${\mathcal N}=2\;$} SQCD, the basis of everything.
In Sec.~\ref{sec:wcp} we discuss two-dimensional WCP(2,2) model, including twisted mass terms. Section \ref{2-4}
is devoted to 2D-4D correspondence. Section \ref{conifold} explains how a massless baryon manifests itself in string theory. Section \ref{sec:kink_mass} is devoted to exact superpotential, vacua of the 2D theory and massive
excitations over the vacua. All relevant central charges are calculated. Then we choose a hierarchy of mass terms in which a part of our analysis can be carried out in terms of the CP(1) model.
In Sec.~\ref{sec:weak} we discuss the weak coupling spectrum while Sec.~\ref{sec:mirror} is devoted to the mirror description of the strong coupling states.
Strong and weak coupling regions are separated by curves of marginal stability which are discussed in Sec.~\ref{sec:CMS}.
After the \textquote{ordinary} spectra are established, we present the non-perturbative Higgs branch and the re-discovered baryon in Sec.~\ref{sec:baryon_2d}.
In Sec.~\ref{sec:2D_4D} we discuss the relation between the bulk and world-sheet theories. 
In the semiclassical approximation the coupling constant $\beta$  in the world-sheet  sigma model is related to the bulk SU(2) gauge coupling $g^2$ via
$$ \beta = \frac{4\pi}{g^2}\,.$$ We derive the exact relation between 2D and 4D couplings in Sec.~\ref{sec:2D_4D} where it is compared with the previous result \cite{Karasik}. We also discuss the strong-weak dualities in the two and four-dimensional theories.
Section \ref{sec:conclusions} summarizes  our conclusions.

%
%

\section {Non-Abelian vortices}
\label{nastrings}
\setcounter{equation}{0}

\subsection{Four-dimensional \boldmath{${\mathcal N}=2\;$} SQCD}
\label{SQCD}

Non-Abelian vortex strings were first found in 4D
\ntwo SQCD with the gauge group U$(N)$ and $N_f \ge N$ quark flavors  
\cite{SYmon,HT2,HT1,ABEKY}, see  \cite{SYrev,Trev,Jrev,Trev2} for   review.
In particular, the matter sector of  the U$(N)$ theory contains
 $N_f$ quark hypermultiplets  each consisting
of   the complex scalar fields
$q^{kA}$ and $\widetilde{q}_{Ak}$ (squarks) and
their  fermion superpartners -- all in the fundamental representation of 
the SU$(N)$ gauge group.
Here $k=1,..., N$ is the color index
while $A$ is the flavor index, $A=1,..., N_f$. We also introduce quark masses $m_A$ as an IR regularization. In the end,
to make contact with string theory, 
we consider the massless limit $m_A\to 0$.
In addition, we introduce the Fayet–Iliopoulos
(FI) parameter $\xi$ in the U(1) factor of the gauge group.
It does not break \ntwo supersymmetry.

At weak coupling, $g^2\ll 1$ (here $g^2$ is the SU$(N)$ gauge coupling), this theory is in the Higgs regime in which squarks develop vacuum 
expectation values (VEVs).
The squark VEV's  are  
\beqn
\langle q^{kA}\rangle &=& \sqrt{\xi}\,
\left(
\begin{array}{cccccc}
1 & \ldots & 0 & 0 & \ldots & 0\\
\ldots & \ldots & \ldots  & \ldots & \ldots & \ldots\\
0 & \ldots & 1 & 0 & \ldots & 0\\
\end{array}
\right), \qquad  \langle\bar{\widetilde{q}}^{kA}\rangle= 0,
\nonumber\\[4mm]
k&=&1,..., N\,,\qquad A=1,...,N_f\, ,
\label{qvev}
\eeqn
where the squark fields are presented as matrices in the color ($k$) and flavor ($A$) indices (the small Latin letters mark the lines in
this matrix while capital letters mark the rows).

These VEVs break 
the U$(N)$ gauge group. As a result, all gauge bosons are
Higgsed. The Higgsed gauge bosons combine with the screened quarks to form 
long \ntwo multiplets, with the mass 
\beq
m_G \sim g\sqrt{\xi}\,.
\label{ximG}
\eeq

In addition to  the U$(N)$ gauge symmetry, the squark condensate (\ref{qvev}) 
breaks also the flavor SU$(N_f)$ symmetry. If the quark masses vanish,
a diagonal global SU$(N)$ combining the gauge SU$(N)$ and an
SU$(N)$ subgroup of the flavor SU$(N_f)$
group survives, however.  This is a well known phenomenon of color-flavor locking. 

Thus, the unbroken global symmetry of our 4D SQCD is  
\beq
 {\rm SU}(N)_{C+F}\times {\rm SU}(\tN)\times {\rm U}(1)_B\,.
\label{c+f}
\eeq
Above,  
$$\tN=N_f-N\,.$$ 
This U(1) in (\ref{c+f}) is associated with $\tN$ quarks with the flavor indices $A= N+1, \, N+2, ..., N_F$, see \cite{SYrev} for more details. 
More exactly,  our U(1)$_B$
is  an unbroken (by the squark VEVs) combination of two U(1) symmetries:  the first is a subgroup of the flavor 
SU$(N_f)$ and the second is the global U(1) subgroup of U$(N)$ gauge symmetry.

 The unbroken global U(1)$_B$ factor in Eq. (\ref{c+f})  is identified with a ``baryonic" symmetry. Note that 
what is usually identified as the baryonic U(1) charge is a {\em part} of  our 4D SQCD  {\em gauge} group.

The 4D theory has a Higgs branch ${\cal H}$ formed by massless quarks which are in  the bifundamental representation
of the global group \eqref{c+f} and carry baryonic charge, see \cite{KSYconifold} for more details.
The dimension of this branch is 
\beq
{\rm dim}\,{\cal H}= 4N \tN.
\label{dimH}
\eeq
This perturbative Higgs branch is an exact property of the theory and can be continued all the way to strong coupling.
 
Below we focus on the particular case $N=2$ and $N_f=4$ because, as was mentioned in Sec.~\ref{sec:introduction}, in this case 4D \ntwo SQCD supports non-Abelian vortex strings which behave as critical superstrings \cite{SYcstring}. In this case the global
group \eqref{c+f} reduces to the one in \eqref{globgroup_d=4}. Also, for $N_f=2N$ the gauge coupling $g^2$ of the 4D SQCD does not run; the $\beta$ function 
vanishes. However, the conformal invariance of the theory is explicitly broken by the FI parameter $\xi$, which 
defines VEV's of quarks, see \eqref{qvev}. The FI parameter is not renormalized either. If we introduce non-zero quark masses the Higgs branch \eqref{dimH} is lifted and 
bifundamental quarks acquire masses $(m_P-m_K)$, $P=1,2$, $K=3,4$. Note that bifundamental quarks form  short BPS multiplets
and their masses do not receive quantum corrections, see \cite{SYrev} for details.

As was already noted, we consider \ntwo SQCD  in the Higgs phase:  $N$ squarks  condense. 
Therefore, the non-Abelian 
vortex strings at hand confine monopoles. In the \ntwo 4D bulk theory the above strings are 1/2 BPS-saturated; hence,  their
tension  is determined  exactly by the FI parameter,
\beq
T=2\pi \xi\,.
\label{ten}
\eeq
However, 
the monopoles cannot be attached to the string endpoints because in U$(N)$ theories strings are topologically stable. In fact, in the U$(N)$ theories confined  
 monopoles 
are  junctions of two distinct elementary non-Abelian strings \cite{SYmon,HT2,T} (see \cite{SYrev} 
for a review). As a result,
in  4D \ntwo SQCD we have 
monopole-antimonopole mesons in which monopole and antimonopole are connected by two confining strings, see Fig.~\ref{monmb}a.
 In addition, in the U$(N)$  gauge theory we can have baryons  appearing as  a closed 
``necklace'' configurations of $N\times$(integer) monopoles \cite{SYrev}. For the U(2) gauge group the 
important example of a baryon consists of four monopoles as shown in Fig.~\ref{monmb}b.

Both stringy monopole-antimonopole mesons and monopole baryons with spins $J\sim 1$ have masses determined 
by the string tension,  $\sim \sqrt{\xi}$ and are heavier at weak coupling $g^2\ll 1$ than perturbative states with masses
$m_G\sim g\sqrt{\xi}$. 
Thus they can decay into perturbative states \footnote{Their quantum numbers with respect to the global group 
\eqref{c+f} allow these decays, see \cite{SYrev}.} and in fact at weak coupling we do not 
expect them to appear as stable  states.

Only in the   strong coupling domain $g^2\sim 1$  we can expect that (at least some of) stringy mesons and baryons
shown in Fig.~\ref{monmb} become stable. We show in this paper that in much the same way as in asymptotically free
versions of \ntwo SQCD (see \cite{SYdualrev}) our 4D theory enters the {\em instead-of-confinement} phase where
quarks and gluons screened  at weak coupling evolve into stringy mesons  as we move to the strong coupling region.

\subsection{World-sheet sigma model}
\label{sec:wcp}

The presence of the color-flavor locked group SU$(N)_{C+F}$ is the reason for the formation of the
non-Abelian vortex strings \cite{SYmon,HT2,HT1,ABEKY}.
The most important feature of these vortices is the presence of the  orientational  zero modes.
As was already mentioned, in \ntwo SQCD these strings are 1/2 BPS saturated. 

Let us briefly review the model emerging on the world sheet
of the non-Abelian  string \cite{SYrev}.

The translational moduli fields  are described by the Nambu–Goto action\,\footnote{In the supersymmetrized form.} and  decouple from all other moduli. Below we focus on
 internal moduli.

If $N_f=N$  the dynamics of the orientational zero modes of the non-Abelian vortex, which become 
orientational moduli fields 
 on the world sheet, are described by two-dimensional (2D)
\ntwot supersymmetric ${\mathbb{CP}}(N-1)$ model.

If one adds additional quark flavors, non-Abelian vortices become semilocal --
they acquire size moduli \cite{AchVas}.  
In particular, for the non-Abelian semilocal vortex in U(2) \ntwo SQCD with four flavors,  in 
addition to  the complex orientational moduli  $n^P$ (here $P=3,4$), we must add the size moduli   
$\rho^K$ (where $K=1,2$), see \cite{HT2,HT1,AchVas,SYsem,Jsem,SVY}. The size moduli are also 
complex.  

The effective theory on the string world sheet is a two-dimensional \ntwot weighted CP sigma model, which we denote
\wcpt \footnote{Both the orientational and the size moduli
have logarithmically divergent norms, see e.g.  \cite{SYsem}. After an appropriate infrared 
regularization, logarithmically divergent norms  can be absorbed into the definition of 
relevant two-dimensional fields  \cite{SYsem}.
In fact, the world-sheet theory on the semilocal non-Abelian string is 
not exactly the \wcp  model \cite{SVY}, there are minor differences. The actual theory is called the $zn$ model. Nevertheless it has the same infrared physics as the model (\ref{wcp22}) \cite{KSVY}, see also \cite{CSSTY}.} %
\cite{SYcstring,KSYconifold,KSYcstring}%
.
This model describes internal dynamics of the non-Abelian semilocal string. 
For details see e.g. the review \cite{SYrev}.

The \wcpt sigma model 
can be  defined  as a low energy limit of the  U(1) gauge theory \cite{W93}.  The bosonic part of the action reads
 \footnote{Equation 
(\ref{wcp22}) and similar expressions below are given in Euclidean notation.}
\begin{equation}
\begin{aligned}
	&S = \int d^2 x \left\{
	\left|\nabla_{\alpha} n^{P}\right|^2 
	+\left|\tilde{\nabla}_{\alpha} \rho^K\right|^2
	+\frac1{4e^2}F^2_{\alpha\beta} + \frac1{e^2}\,
	\left|\pt_{\alpha}\sigma\right|^2
	\right.
	\\[3mm]
	&+\left.
	2\left|\sigma+\frac{m_P}{\sqrt{2}}\right|^2 \left|n^{P}\right|^2 
	+ 2\left|\sigma+\frac{m_{K}}{\sqrt{2}}\right|^2\left|\rho^K\right|^2
	+ \frac{e^2}{2} \left(|n^{P}|^2-|\rho^K|^2 - r \right)^2
	\right\},
	\\[4mm]
	&
	P=1,2\,,\qquad K=3,4\,.
\end{aligned}
\label{wcp22}
\end{equation}
Here, $m_A$ ($A=1,..,4$) are the so-called twisted masses (they come from 4D quark masses),
while $r$ is the inverse coupling constant (2D FI term). Note that $r$ is the real part of the complexified coupling constant 
introduced in Eq. (\ref{beta_complexified}), $$r = {\rm Re}\,\beta\,.$$

The fields $n^{P}$ and $\rho^K$ have
charges  $+1$ and $-1$ with respect to the auxiliary U(1) gauge field, and the corresponding  covariant derivatives in (\ref{wcp22}) are defined as 
\begin{equation}
	\nabla_{\alpha}=\p_{\alpha}-iA_{\alpha}\,,
	\qquad 
	\tilde{\nabla}_{\alpha}=\p_{\alpha}+iA_{\alpha}\,,	
\end{equation}
respectively. The complex scalar field $\sigma$ is a superpartner of the U(1) gauge field $A_{\alpha}$.

The number of real bosonic degrees of freedom in the model \eqref{wcp22} is $8-1-1=6$.  Here 8 is the number of real degrees of 
freedom of $n^P$ and $\rho^K$ fields and we subtracted one real constraint imposed by the last  term in \eqref{wcp22} in the limit $e^2\to \infty$  and one  gauge phase eaten by the Higgs mechanism.

Apart from the U(1) gauge symmetry, the sigma model (\ref{wcp22}) in the massless limit has a global symmetry group
\begin{equation}
	 {\rm SU}(2)\times {\rm SU}(2)\times {\rm U}(1)_B \,,
\label{globgroup}
\end{equation}
i.e. exactly the same as the unbroken global group in the 4D theory at $N=2$ and $N_f=4$ \eqref{globgroup_d=4}. 
The fields $n$ and $\rho$ 
transform in the following representations:
\begin{equation}
	n:\quad \left(\textbf{2},\,\textbf{1},\, 0\right), \qquad \rho:\quad \left(\textbf{1},\,\textbf{2},\, 1\right)\,.
\label{repsnrho}
\end{equation}
This his been already presented in (\ref{ksat}).
Here  the global ``baryonic''  U(1)$_B$ symmetry is a classically unbroken (at $\beta >0$) combination of the 
global U(1) group which
rotates $n$ and $\rho$ fields with the same phases plus  U(1) gauge symmetry which  rotates them with the opposite phases, see
\cite{KSYconifold} for details.
Non-zero twisted masses $m_A$ break each of the SU(2) factors in \eqref{globgroup} down to U(1).

The 2D coupling constant $r$ can be naturally complexified if we
include the $\theta$ term in the action,
\begin{equation}
	\beta = r + i \, \frac{\theta_{2d}}{2 \pi} \,,
\label{beta_complexified}	
\end{equation}
where $\theta_{2d}$ is the two-dimensional $\theta$ angle. 

At  the quantum level, the coupling $\beta$ does not run in this theory. Thus, 
the \wcpt  model is superconformal at zero masses $m_A \equiv 0$. The model \eqref{wcp22} is a mass deformation of this superconformal theory.

From action \eqref{wcp22} for \wcpt it is obvious that this model is self-dual. The duality transformation
\begin{equation}
\begin{aligned}
	\beta &\to \wt{\beta} = - \beta \, \\
	m_{1,2} &\to \wt{m}_{1,2} = - m_{3,4} \, \\
	m_{3,4} &\to \wt{m}_{3,4} = - m_{1,2} \, \\
	\sigma &\to \wt{\sigma} = - \sigma
\end{aligned}
\label{2d_S-duality}
\end{equation}
exchanges the roles of the orientation moduli $n^P$ and size moduli $\rho^K$. The point $\beta = 0$ is the {\em self-dual} point.

\subsection{2D-4D correspondence}
\label{2-4}

As was mentioned above confined monopoles of 4D SQCD are junctions of two different elementary non-Abelian strings. In  the  world-sheet theory they are seen as kinks interpolating between different vacua of
$\mathbb{WCP}(N,\tN)$ model. This ensures 2D-4D correspondence: the coincidence between the BPS spectrum of 
monopoles in 4D SQCD at a particular singular point on the Coulomb branch (which becomes  the quark vacuum \eqref{qvev} once we introduce non-zero $\xi$)  and the spectrum of kinks in 2D $\mathbb{WCP}(N,\tN)$ model. The masses of (dyonic) monopoles in  4D SQCD are 
given by the exact Seiberg-Witten solution \cite{SW2}, while the kink spectrum in $\mathbb{WCP}(N,\tN)$ model can be 
derived from exact twisted effective superpotential  \cite{Dorey,W93,AdDVecSal,ChVa,HaHo,DoHoTo}. This effective superpotential is written in terms of the twisted chiral superfield which has the complex scalar  field $\sigma$ 
(see \eqref{wcp22})
as its lowest component \cite{W93}, see Sec.~\ref{sec:kink_mass} where we introduce this superpotential and study
the kink spectrum for the \wcpt model.

This coincidence was observed in \cite{Dorey,DoHoTo} and   explained later 
in \cite{SYmon,HT2} using the picture of confined bulk monopoles which are seen as kinks in the world 
sheet theory. A crucial point is that both the monopoles and the kinks are BPS-saturated states\,\footnote{Confined
 monopoles, being junctions of two distinct 1/2-BPS strings, are 1/4-BPS states in 4D SQCD 
\cite{SYmon}.},
and their masses cannot depend on the non-holomorphic parameter $\xi$ \cite{SYmon,HT2}. This means that,
although the confined monopoles look physically very different from unconfined monopoles on the Coulomb branch
of 4D SQCD (in a particular singular point that becomes the isolated vacuum at nonzero $\xi$),
their masses are the same. Moreover, these masses coincide with the masses of kinks in the world-sheet 
theory.

Note that  VEVs of $\sigma$ given by the exact twisted superpotential  coincide
with the double roots of the Seiberg-Witten curve \cite{SW2} in the quark vacuum of
4D SQCD \cite{Dorey,DoHoTo}. This is the key technical reason that leads to the
coincidence of the  2D and 4D BPS spectra.

%
%

\section {Massless 4D baryon from  string theory}
\label{conifold}
 \setcounter{equation}{0}

 The world-sheet \wcpt model \eqref{wcp22} is conformal and due to \ntwot supersymmetry the metric of its target space is 
K\"ahler. The conformal invariance of the model also ensures that this metric is Ricci flat. Thus the target space of model \eqref{wcp22} is a Calabi-Yau manifold. 

Moreover, as we explained in the previous subsection the world-sheet \wcpt \\
model has six real bosonic degrees of freedom.
Its target space defined by the  $D$-term condition
\beq
|n^P|^2-|\rho^K|^2 = \beta\,
\label{Fterm}
\eeq 
is a six dimensional non-compact Calabi-Yau space $Y_6$ known as  conifold, see \cite{NVafa} for a review. Together with
four translational moduli of the non-Abelian vortex it forms a ten dimensional target space $\mathbb{R}^4\times Y_6$ required for 
a superstring to be critical \cite{SYcstring}. 

In this section we briefly review the only 4D massless state found from the string theory of the critical non-Abelian vortex \cite{KSYconifold}. It is associated 
with the deformation of the conifold complex structure. 
 As was already mentioned, all other massless string modes  have non-normalizable wave functions over the conifold. In particular, 4D graviton associated with a constant wave
function over the conifold $Y_6$ is
absent \cite{KSYconifold}. This result matches our expectations since we started with
\ntwo SQCD in the flat four-dimensional space without gravity.

We can construct the U(1) gauge-invariant ``mesonic'' variables
\beq
w^{PK}= n^P \rho^K.
\label{w}
\eeq
These variables are subject to the constraint
\beq
{\rm det}\, w^{PK} =0. 
\label{coni}
\eeq

Equation (\ref{coni}) defines the conifold $Y_6$.  
It has the K\"ahler Ricci-flat metric and represents a non-compact
 Calabi-Yau manifold \cite{W93,NVafa,Candel}. It is a cone which can be parametrized 
by the non-compact radial coordinate 
\beq
\widetilde{r}^{\, 2} = {\rm Tr}\, \bar{w}w\,
\label{tilder}
\eeq
and five angles, see \cite{Candel}. Its section at fixed $\widetilde{r}$ is $S_2\times S_3$.

At $\beta =0$ the conifold develops a conical singularity, so both $S_2$ and $S_3$  
can shrink to zero.
The conifold singularity can be smoothed out
in two distinct ways: by deforming the K\"ahler form or by  deforming the 
complex structure. The first option is called the resolved conifold and amounts to keeping
a non-zero value of $\beta$ in (\ref{Fterm}). This resolution preserves 
the K\"ahler structure and Ricci-flatness of the metric. 
If we put $\rho^K=0$ in (\ref{wcp22}) we get the $\mathbb{CP}(1)$ model with the $S_2$ target space
(with the radius $\sqrt{\beta}$).  
The resolved conifold has no normalizable zero modes. 
In particular, 
the modulus $\beta$  which becomes a scalar field in four dimensions
 has non-normalizable wave function over the 
$Y_6$ and therefore is not dynamical \cite{KSYconifold}.  

If $\beta=0$ another option exists, namely a deformation 
of the complex structure \cite{NVafa}. 
It   preserves the
K\"ahler  structure and Ricci-flatness  of the conifold and is 
usually referred to as the {\em deformed conifold}. 
It  is defined by deformation of Eq.~(\ref{coni}), namely,   
\beq
 {\rm det}\, w^{PK} = b\,,
\label{deformedconi}
\eeq
where $b$ is a complex number.
Now  the $S_3$ can not shrink to zero, its minimal size is 
determined by
$b$. 

The modulus $b$ becomes a 4D complex scalar field. The  effective action for  this field was calculated in \cite{KSYconifold}
using the explicit metric on the deformed conifold  \cite{Candel,Ohta,KlebStrass},
\beq
S(b) = T\int d^4x |\pt_{\mu} b|^2 \,
\log{\frac{T^2 L^4}{|b|}}\,,
\label{Sb}
\eeq
where $L$ is the  size of $\mathbb{R}^4$ introduced as an infrared regularization of 
logarithmically divergent $b$ field 
norm.\footnote{The infrared regularization
on the conifold $\widetilde{r}_{\rm max}$ translates into the size $L$ of the 4D space 
 because the variables  $\rho$ in \eqref{tilder} have an interpretation of the vortex string sizes,
$\widetilde{r}_{\rm max}\sim TL^2$ .}

We see that the norm of
the $b$ modulus turns out to be  logarithmically divergent in the infrared.
The modes with the logarithmically divergent norm are at the borderline between normalizable 
and non-normalizable modes. Usually
such states are considered as ``localized'' ones. We follow this rule. 
This scalar mode is localized near the conifold singularity  in the same sense as the orientational 
and size zero modes are localized on the vortex-string solution.
   
 The field $b$  being massless can develop a VEV. Thus, 
we have a new Higgs branch in 4D \ntwo SQCD which is developed only for the critical value of 
the 4D coupling constant \footnote{The complexified 4D coupling constant $\tau=1$ at this point, see Sec.~\ref{sec:2D_4D}.} associated with $\beta=0$.

 In \cite{KSYconifold} the massless state $b$ was interpreted as a baryon of 4D \ntwo QCD.
Let us explain this.
 From Eq.~(\ref{deformedconi}) we see that the complex 
parameter $b$ (which is promoted to a 4D scalar field) is a singlet with respect to both SU(2) factors in
 (\ref{globgroup}), i.e. 
the global world-sheet group.\footnote{Which is isomorphic to the 4D
global group \eqref{globgroup_d=4} .} What about its baryonic charge? From \eqref{repsnrho} and \eqref{deformedconi}
we see that the $b$ state transforms as 
\beq
({\bf 1},\,{\bf 1},\,2).
\label{brep}
\eeq
 In particular it has the baryon charge $Q_B(b)=2$.

To conclude this section let us note that in type IIA superstring the complex scalar 
associated with deformations of the complex structure of the Calabi-Yau
space enters as a 4D \ntwo BPS hypermultiplet. Other components of this hypermultiplet can be restored 
by \ntwo supersymmetry. In particular, 4D \ntwo hypermultiplet should contain another complex scalar $\tilde{b}$
with baryon charge  $Q_B(\tilde{b})=-2$. In the stringy description this scalar comes from ten-dimensional
three-form, see \cite{Louis} for a review.

Below in this paper we study the BPS kink spectrum of the world-sheet model \eqref{wcp22} using purely field theory
methods. Besides other results we use the  2D-4D correspondence to confirm  the emergence of 4D baryon with quantum numbers
\eqref{brep} and the presence of the associated non-perturbative Higgs branch at $\beta=0$.

%
%

\section{Kink mass from the exact superpotential \label{sec:kink_mass}}

As was mentioned above, the \wcpt model \eqref{wcp22} supports BPS saturated kinks interpolating between different vacua. 
In this section we will obtain the kink central charges and, consequently, their masses.

\subsection{Exact central charge}

For the model at hand we can obtain an exact formula for the BPS kink central charge. 
This is possible because for this model an exact twisted superpotential obtained by integrating out $n$ and $\rho$ supermultiplets  is known. It is a generalization \cite{HaHo,DoHoTo}
of the CP($N-1$) model superpotential \cite{Dorey,W93,AdDVecSal,ChVa} of the  Veneziano-Yankielowicz  type \cite{VYan}.
In the present case $N_f = 2N = 4$ it reads:
\begin{multline}
	 {\cal W}_{\rm WCP}(\sigma)= \frac{1}{4\pi}\Bigg\{ 
	 	\sum_{P=1,2} \left( \sqrt{2} \, \sigma + m_P \right) \ln\left( \sqrt{2} \, \sigma + m_P \right)
	 	\\
	 	- \sum_{K=3,4} \left( \sqrt{2} \, \sigma + m_K \right) \ln\left( \sqrt{2} \, \sigma + m_K \right)
	 	+ 2 \pi \,  \sqrt{2} \, \sigma  \, \beta
	 	+ \text{const}
	 \Bigg\}\,,
\label{WCPsup}
\end{multline}
where we use one and the same notation $\sigma$ for the  twisted superfield \cite{W93} and its lowest scalar
component. 
To study the vacuum structure of the theory we minimize this superpotential with respect to $\sigma$ to obtain the 2D vacuum equation
\begin{equation}
	\prod_{P=1,2}\left(\sqrt{2} \, \sigma + m_P \right) 
		= e^{- 2 \pi \beta} \cdot \prod_{K = 3,4} \left(\sqrt{2} \, \sigma + m_K \right) \,.
\label{2d_equation}	
\end{equation}
The invariance of equation \eqref{2d_equation}	under the duality transformation \eqref{2d_S-duality} is evident. 

The vacuum equation \eqref{2d_equation} has two solutions (VEVs) $\sigma_{1,2}$, which means that generically there are two degenerate vacua in our theory. 
Therefore, there are BPS kinks interpolating between these two vacua.
Their masses are given by the absolute value of the central charge, 
\beq
M_{\rm BPS} = |Z|. 
\label{MZ}
\eeq
The central charge can be found by taking the appropriate difference of the superpotential \eqref{WCPsup} calculated at distinct roots \cite{Dorey,HaHo,DoHoTo}.
Say, for the kink interpolating between the vacua $\sigma_2$ and $\sigma_1$, the central charge is given by
\begin{equation}
\begin{aligned}
	Z_{\rm BPS} 
		&= 2\left[{\cal W}_{\rm WCP}(\sigma_{1})-{\cal W}_{\rm WCP}(\sigma_{2})\right] 	\\
		&= \frac{1}{2\pi}\Bigg[ 
			\sum_{P=1}^{N_c} m_P \ln\frac{\sqrt{2}\sigma_1 + m_P}{\sqrt{2}\sigma_2 + m_P}
			- \sum_{K=N_c+1}^{N_f} m_K \ln\frac{\sqrt{2}\sigma_1 + m_K}{\sqrt{2}\sigma_2 + m_K}
		\Bigg]
		\,.
\end{aligned}
\label{BPSmass}
\end{equation}
Note that in order for this equation to transform well under the $S$ duality transformation, we must assume that the masses are transformed as $m_P \to - m_K$, $m_K \to - m_P$.

The central charge formula \eqref{BPSmass} contains logarithms, which are multivalued. Distinct choices differs by contributions $im_A \times {\rm integer}$.  In addition to the
topological charge, the kinks can carry Noether charges with respect to the global group \eqref{globgroup} broken down to
U(1)$^3$ by the mass differences. This produces a whole family of dyonic kinks. We stress that all these kinks  interpolate
between the same pair of vacua $\sigma_1$ and $\sigma_2$. In Eq.~\eqref{BPSmass} we do not specify these dyonic contributions. 
Below in this paper we present a detail study of the BPS kink spectrum in different regions of the coupling constant $\beta$.

The $\sigma$ vacua are found by solving equation \eqref{2d_equation},
\begin{equation}
	\sqrt{2} \sigma_{\pm} = - \frac{\Delta m}{2} \, \frac{1 + e^{- 2 \pi \beta}}{1 - e^{- 2 \pi \beta}} 
		~\pm~ \sqrt{ \frac{(\delta m_{12})^2 - e^{- 2 \pi \beta} \, (\delta m_{34})^2}{4 (1 - e^{- 2 \pi \beta}) } + \Delta m^2 \, \frac{e^{- 2 \pi \beta}}{(1 - e^{- 2 \pi \beta})^2}}
		\,.
\label{roots_symmetric}	
\end{equation}
In writing down this formula we have used the following parametrization of the masses:
\begin{equation}
\begin{aligned}
	\Delta m      &~=~  \ov{m} - \wt{m} \,,\quad { \Delta m^2      ~=~  \big(\ov{m} - \wt{m}\big)^2\,,}\\
	\delta m_{12} &~=~ m_1 - m_2 \,, \\
	\delta m_{34} &~=~ m_3 - m_4 \,,
\end{aligned}
\label{mass_parametrization}
\end{equation}
where $\ov{m}$ and $\wt{m}$ are the averages of bare masses of the $n^{P}$ and $\rho^{K}$ fields, respectively, 
\begin{equation}
	\ov{m} = \frac{m_1 + m_2}{2} 
	\,, \quad
	\wt{m} = \frac{m_3 + m_4}{2} \,.
\label{mbar_mtilde}
\end{equation}
From \eqref{roots_symmetric} we immediately observe that generically one of the roots grows indefinitely near the self-dual point $\beta=0$, while the other remains finite. This will turn out to be important for consideration of kinks at strong coupling.

The Argyres-Douglas (AD) points \cite{AD} correspond to
fusing the two vacua. In these points certain kinks become massless.  Given the solution \eqref{roots_symmetric}, the AD points arise when  the expression under the square root 
vanishes. 
The formula for the positions of the AD points in the $\beta$ plane can be expressed as
\begin{equation}
	e^{- 2 \pi \beta_{AD}} =  \big( 2 P - 1 \pm 2 \sqrt{P (P - 1)} \big) \cdot \frac{m_1 - m_2}{m_3 - m_4} \,,
\label{AD_general}
\end{equation}
where $P$ is a conformal cross-ratio,
\begin{equation}
	P[m_1, m_4, m_3, m_2] 
		= \frac{(m_1 - m_4) (m_3 - m_2)}{(m_1 - m_2) (m_3 - m_4)} \,.
\end{equation}
Formula \eqref{AD_general} may have singularities. Values $P=0,1$ correspond to a Higgs brunch opening up, while at $P\to\pm\infty$ one of AD points runs away to $\beta\to\pm\infty$. There is not much interesting going on at these singularities, and at generic masses formula \eqref{AD_general} is perfectly fine. Therefore, we will not consider  these points here.

\subsection{CP(1) limit}

To make contact with the well understood kink spectrum of $\mathbb{CP}(1)\;$ model 
 we consider the following limit\footnote{In this section and below similar inequalities involving $\gg$ or $\ll$ we actually assume that on the l.h.s. we take the absolute value of masses and real part of $\beta$, e.g. \eqref{CP1_limit} actually means $| \Delta m | \gg | \delta m_{12}|  \,, \ | \delta m_{34} | $; $\Re\beta \gg 1$.}:
\begin{equation}
	\Delta m \gg \delta m_{12} \,, \ \delta m_{34}
	\,;	\quad 
	\beta \gg 1 \,.
\label{CP1_limit}
\end{equation}
Most of the general features (with the exception of the weak coupling bound states, see Sec.~\ref{sec:weak}) of the WCP(2,2) model are still preserved in this limit, but calculations simplify greatly.
Moreover, results of this section easily generalize to the case $\beta \ll -1$.

By an appropriate redefinition of the $\sigma$ field we can shift the masses to
\begin{equation}
\begin{aligned}
	m_1 &= \delta m_{12} / 2 \,, \quad
	m_2 = - \delta m_{12} / 2 \,, \\
	m_3 &= - \Delta m + \delta m_{34} / 2 \,, \quad
	m_4 = - \Delta m - \delta m_{34} / 2 \,.
\end{aligned}
\label{CP1_masses}
\end{equation}
In this representation it is evident that in the limit \eqref{CP1_limit} the $\rho^K$ fields are heavy and decouple at energies below $\Delta m$, and the theory at low energies reduces to the ordinary $\mathbb{CP}(1)\;$ model with mass scale $\delta m_{12}$. 
The effective coupling constant is no longer constant. It runs  below $\Delta m$ and freezes at the scale $\delta m_{12}$,
\beq
2\pi\beta_{CP(1)} = 2\pi \beta - 2 \ln \frac{\Delta m }{\delta m_{12}}= 2 \log{\left\{\frac{\delta m_{12}}{\Lambda_{CP(1)}}\right\}},
\label{CP1_beta}
\eeq
where the factor 2 in the r.h.s. is the first coefficient of the $\beta$ function (for  $\mathbb{CP}(N-1)\;$ this coefficient is 
$N$),
while $\Lambda_{CP(1)}$ is the dynamical scale of the low-energy $\mathbb{CP}(1)\;$ model,
\begin{equation}
	\Lambda_{CP(1)} = \Delta m \, e^{- \pi \beta} \,.
\label{CP1_scale}	
\end{equation}

The vacuum equation \eqref{2d_equation} becomes 
\begin{equation}
	(\sqrt{2} \sigma - \delta m_{12} / 2) \, (\sqrt{2} \sigma + \delta m_{12} / 2) \approx 
	e^{- 2 \pi \beta} (- \Delta m)^2 \, = \Lambda_{CP(1)}^2.
	\label{k414}
\end{equation}
In the limit \eqref{CP1_limit},  Eq. (\ref{k414})  fits the $\mathbb{CP}(N-1)\;$ vacuum equation
\begin{equation}
	\prod_{P=1}^{N}\left(\sqrt{2}\sigma + m_P \right) = \left( \Lambda_{CP(N-1)} \right)^{N}
\end{equation}
(for $N=2$). 
In the limit \eqref{CP1_limit} the AD points \eqref{AD_general} are given by $\pm \beta_{AD}$ with
\begin{equation}
	\beta_{AD} \approx \frac{1}{\pi} \ln\frac{2 \Delta m}{\delta m_{12}} \pm \frac{i}{2}  \,.
\label{AD_roots_simplecase_beta1_approx}
\end{equation}
We see that the $\mathbb{CP}(1)\;$ weak coupling condition $\Lambda_{CP(1)} \ll \delta m_{12}$ directly translates to $\beta \gg \beta_{AD}$,
see \eqref{CP1_scale}. Let us stress that this  is more restrictive condition then just $\beta \gg 1$. If $\beta \to \beta_{AD}$
the effective coupling \eqref{CP1_beta} hits the infrared pole.

Now, since we are in the $\mathbb{CP}(1)\;$ limit, the BPS kink central charge must be given by the well known formula \cite{Dorey}. Indeed, the 2D vacua are approximately given by
\begin{equation}
	\sqrt{2}\sigma_\pm  \approx \pm \frac{1}{2} \sqrt{\delta m_{12}^2 + 4 \, \Lambda_{CP(1)}^2} \,.
\end{equation}
Substituting this and \eqref{CP1_masses} into the \wcpt central charge formula \eqref{BPSmass} and neglecting terms $\delta m_{12} / \Delta m$ and $\Lambda_{CP(1)} / \Delta m$
we obtain for the central charge
\begin{equation}
	Z_\text{kink} = \frac{1}{2\pi}\left[ 
		2 \sqrt{\delta m_{12}^2 + 4 \, \Lambda_{CP(1)}^2}
		- \delta m_{12} \, \ln\frac{\delta m_{12} + \sqrt{\delta m_{12}^2 + 4 \, \Lambda_{CP(1)}^2}}{\delta m_{12} - \sqrt{\delta m_{12}^2 + 4 \, \Lambda_{CP(1)}^2}}
		\right]
		\,.
\label{CP1_Z}
\end{equation}
This is exactly Dorey's  formula \cite{Dorey} for $\mathbb{CP}(1)$.

The above central charge \eqref{CP1_Z} tends to zero at the AD point \eqref{AD_roots_simplecase_beta1_approx}. This ensures that 
the BPS kink becomes massless at this point. We will see later that at two AD points $\beta = \beta_{AD}$ with ${\rm Re}\, \beta >0$ two kinks with distinct
dyonic charges become massless. 

Moreover, the central charge \eqref{CP1_Z} has a singularity at the AD point. Indeed, near this point we have $\delta m_{12}^2 + 4 \Lambda_{CP(1)}^2 \approx 0$. Expanding \eqref{CP1_Z} we get
\begin{equation}
\begin{aligned}
	Z_\text{kink} 
		&\approx - \frac{1}{3\pi \, \delta m_{12}^2} \Big( \delta m_{12}^2 + 4 \,\Lambda_{CP(1)}^2 \Big)^{3/2} \\
		&\approx - \frac{2 \sqrt{2\pi}}{3} \, \delta m_{12} \cdot (\beta - \beta_{AD1})^{3/2} \,.
\label{CP1_Z_AD_2}
\end{aligned}
\end{equation}
This shows that locally the central charge has a root-like singularity near the AD point.

In the quasiclassical limit $\Lambda_{CP(1)} \ll \delta m_{12}$ (or, equivalently, $\beta \gg \beta_{AD}$) the central charge \eqref{CP1_Z} is
\begin{equation}
\begin{aligned}
	Z_\text{kink} 
		&\approx - \frac{\delta m_{12}}{\pi} \, \ln\frac{\delta m_{12}}{\Lambda_{CP(1)}} + i \, \frac{\delta m_{12}}{2} + \frac{\delta m_{12}}{\pi} \\
		&\approx -  \beta_{CP(1)} \cdot (m_1 - m_2) + i \, (m_1 - \ov{m})  + \frac{m_1 - m_2}{\pi} 
\label{CP1_Z_quasiclassical}
\end{aligned}
\end{equation}
where $\ov{m}$ is the average of the first two masses, see \eqref{mbar_mtilde}.
The second term represents the fractional U(1) charge of the soliton \cite{ShVanZwi}. 
Indeed \eqref{CP1_Z_quasiclassical} can be compared to the Dorey quasiclassical formula \cite{Dorey} for the central charge
\begin{equation}
	Z_\text{kink} = -( \beta_{CP(1)} \, T - i  \, q ) \, \delta m_{12}
\label{dorey_quasiclassical}
\end{equation}
where $T=+1$ is the topological charge, while $q$ is the kink global (or \textquote{dyonic}) charge. Comparing \eqref{CP1_Z_quasiclassical} and \eqref{dorey_quasiclassical} we see that the kink dyonic charge is $q = 1/2$. The last term in \eqref{CP1_Z_quasiclassical} is the central charge anomaly \cite{ShVanZwi}. For details see e.g. \cite{SYrev,ShifmanLectureTop}.

In the limit when we are far from any AD point, $\beta \gg \beta_{AD}$, the expression in the second line in \eqref{CP1_Z_quasiclassical} is actually valid for any mass parameters, not just in the $\mathbb{CP}(1)\;$ limit \eqref{CP1_limit}.

%
%

\section{Weak coupling spectrum \label{sec:weak}}

Now let us discuss the weak coupling spectrum. 
In the $\mathbb{CP}(1)\;$ limit \eqref{CP1_limit} at weak coupling, $\Lambda_{CP(1)} \ll \delta m_{12}$, a part of the spectrum coincides with the ordinary $\mathbb{CP}(1)\;$ model spectrum coming from the $n^P$ fields. 

The $\mathbb{CP}(1)\;$ spectrum \cite{Dorey} consists of elementary perturbative excitations and a tower of BPS dyonic kinks.
The perturbative states have a mass $|i (m_1 - m_2) |$. This can be understood on the classical level from the action 
\eqref{wcp22}. Suppose that field $n^1$ classically develops VEV equal to $\sqrt{\beta}$. Then  the first term with $P=1$ in the second line in \eqref{wcp22} forces $\sigma$ to acquire the classical value $\sqrt{2}\sigma =-m_1$, while the term with $P=2$ gives  the mass $|m_1-m_2|$ to $n^2$. Note, that this result obtained in the quasiclassical limit is in fact exact because of the 
BPS nature of this perturbative state.

The mass of a kink interpolating between the two vacua is $M_\text{kink} = |Z_\text{kink}|$ where the central charge is given by 
 \eqref{CP1_Z_quasiclassical}.
This kink is in fact a part of a dyonic tower with central charges
\begin{equation}
	D^{(n)} = Z_\text{kink} + n \cdot i (m_1 - m_2)
	\,, \quad
	n \in \mathbb{Z} \,.
\label{M2_tower}
\end{equation}
which can be interpreted as a bound state of the kink and $n$ quanta of perturbative states with the central charge 
 $i (m_1 - m_2)$.  
The number $n$ in \eqref{M2_tower} is a manifestation of the multiple logarithm brunches in \eqref{BPSmass}. 
It gives a contribution to the kink dyonic charge $q$, see the quasiclassical expression \eqref{dorey_quasiclassical}. 
The total dyonic  charge $q = n + 1/2$ also has a contribution coming from $Z_\text{kink}$ which makes  it non-integer. 
The presence of the tower \eqref{M2_tower} in the weak coupling region of $\mathbb{CP}(1)\;$ model was found in \cite{Dorey} using quasiclassical methods.

In our \wcpt model extra states are present too, coming from the $\rho^K$ fields. 
They include perturbative BPS states with masses $|i (m_P - m_K) |,\ P=1,2,\ K=3,4$. These states are seen at the classical level from the action \eqref{wcp22}. Say, in the classical vacuum $n^1=\sqrt{\beta}$, $\sqrt{2}\sigma =-m_1$, the  fields $\rho^K$
acquire masses $|m_1-m_3|$ and $|m_1-m_4|$ given by the second term in the second line in \eqref{wcp22}.
We will call these states \textquote{bifundamentals.} They are 2D ``images'' of bifundamental quarks of 4D SQCD upon 2D-4D
correspondence, see Sec.~\ref{SQCD}.

If we relax  the $\mathbb{CP}(1)\;$ conditions  \eqref{CP1_limit}, the spectrum described above stays intact.
However we  get some extra states.
States from the dyonic tower \eqref{M2_tower} might form bound states with
\textquote{bifundamental} fermions $\wt{\psi}^{P}_K$. 
The central charge of the resulting state is given by \cite{DoHoTo}
\begin{equation}
	Z^{(n)}_\text{bound} = Z_\text{kink} + n \cdot i (m_1 - m_2) + i (m_1 - m_K) \,.
\label{Z_bound_state}
\end{equation}
These states are formed if the condition
\begin{equation}
	0 < \Re{\frac{m_1 - m_K}{m_1 - m_2}} \equiv 1 - \Re{\frac{m_2 - m_K}{m_2 - m_1}} < 1
\label{bound_stability_condition}	
\end{equation}
is satisfied for some $K \in \{3,4\}$ (and any $P$) \cite{DoHoTo}.
From the stability condition \eqref{bound_stability_condition} it is evident that there are no such bound states 
for our choice of quark masses, $\Delta m \gg \delta m_{12},\,\delta m_{34}$, see the first condition in  \eqref{CP1_limit}.
We will not consider these bound states here.

Here we have just described the spectrum at weak coupling  $\beta \gg \beta_{AD}$. It literally translates into the spectrum in the dual weak coupling region at $\beta \ll - \beta_{AD}$ with substitution of indices $P=1,2 \leftrightarrow K=3,4$.

%
%

\section{Mirror description and the strong coupling spectrum \label{sec:mirror}}

\begin{figure}
	\centering
	\includegraphics[width=0.7\linewidth]{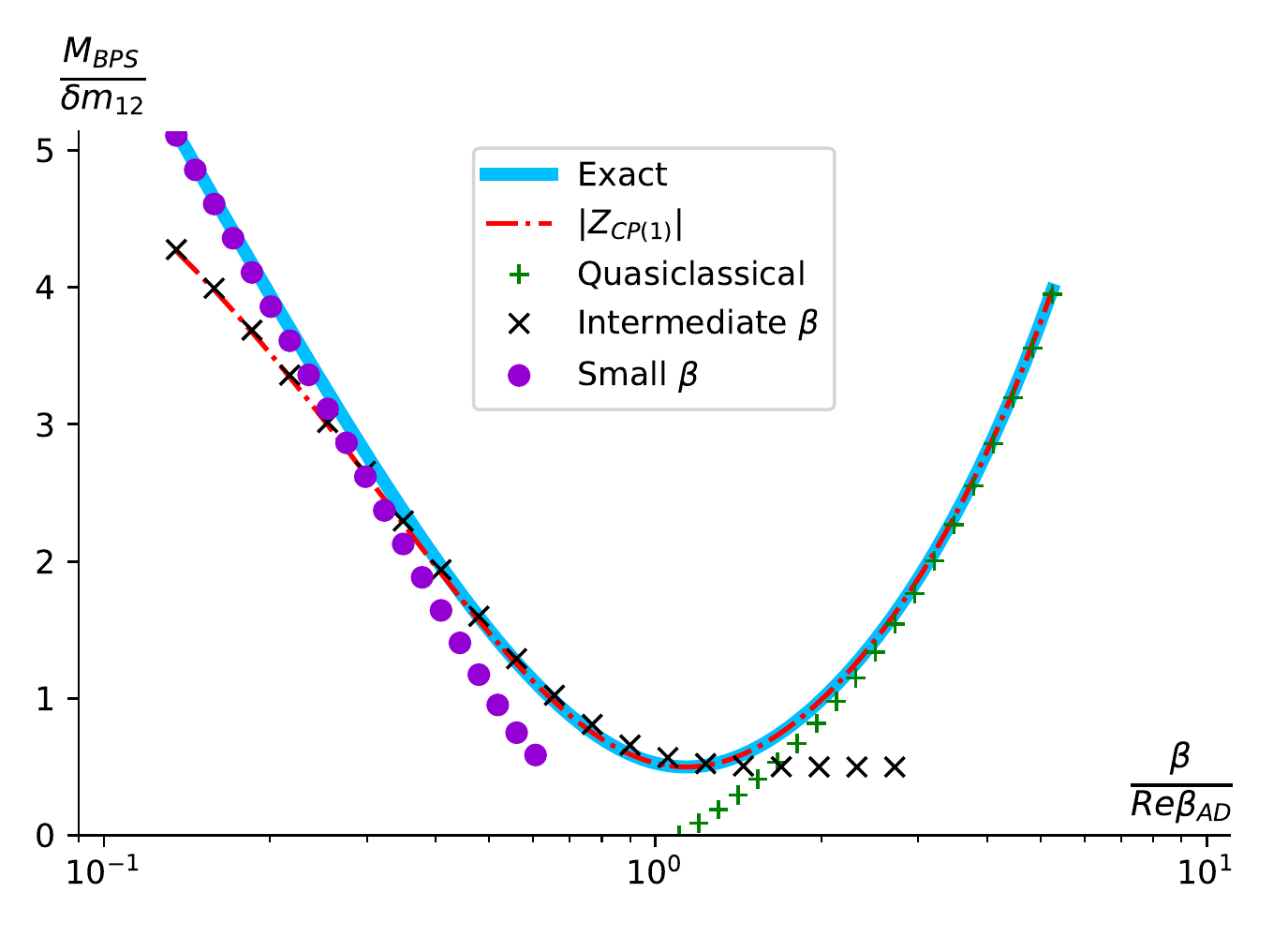}
	\caption{
		Various approximations of the kink mass $M_1$ (absolute value of the central charge): 
		$\mathbb{CP}(1)\;$ limit \eqref{CP1_Z}, 
		quasiclassical \eqref{CP1_Z_quasiclassical}, 
		intermediate $\beta$ \eqref{kink_mass_P_CP1_1},
		small $\beta$ \eqref{kink_mass_P_beta0}.
		Fixed $\Delta m / \delta m_{12} = 10$.
	}
\label{fig:M_approximations}
\end{figure}

In this section we will investigate the BPS kink spectrum in the strong coupling domain where $\beta$ is small, $\beta \ll \beta_{AD}$. 
(For comparison of various limits of the kink mass, see Fig.~\ref{fig:M_approximations}.)
We will generalize the analysis of  \cite{Shifman:2010id} carried out for asymptotically free \wcp models to the present case of 
conformal \wcpt model.

\subsection{Mirror superpotential}

To this end we will implement the mirror description of kinks \cite{FFS,HoVa} of the \wcpt model \eqref{wcp22}. 
The formula for the mirror superpotential is
\begin{equation}
	{\cal W}_{\rm mirror}(X_P, Y_K)
		= - \frac{\Lambda}{4 \pi} \left[ \sum_P X_P - \sum_K Y_K - \sum_P \frac{m_P}{\Lambda} \ln X_P + \sum_K \frac{m_K}{\Lambda} \ln Y_K \right] \,.
\label{W_mirror}		
\end{equation}
Here, the indices run as $P = 1, 2$, $K = 3, 4$. Parameter $\Lambda$ is an auxiliary parameter of dimension of mass which will cancel in the very end. 

The fields $X_P, Y_K$ are subject to the  constraint
\begin{equation}
	\prod_P X_P = e^{- 2 \pi \beta} \prod_K Y_K \,.
\label{mirror_constraint}	
\end{equation}
The VEVs  of $X_P, Y_K$ can be obtained by minimizing the superpotential \eqref{W_mirror} and using the above constraint
\cite{Shifman:2010id,HoVa}. Below we use a simplified approach which utilizes the relation of $X_P, Y_K$ to the $\sigma$ solutions of the vacuum equation \eqref{2d_equation} \cite{Shifman:2010id,HoVa},
\begin{equation}
	X_P = \frac{\sqrt{2}\sigma + m_P}{\Lambda}
	\,, \quad 
	Y_K = \frac{\sqrt{2}\sigma + m_K}{\Lambda} \,.
\label{mirror-x_map}	
\end{equation}

For a kink interpolating between two vacua $Vac_1$ and $Vac_2$,  the central charge is given by an exact formula
\begin{equation}
	Z_\text{kink} = 2 \left[ {\cal W}_{\rm mirror}(Vac_2) - {\cal W}_{\rm mirror}(Vac_1) \right] \,,
\label{mirror_kink_mass}	
\end{equation}
while its mass $M_\text{kink} =|Z_\text{kink}|$, see \eqref{MZ}.

\subsection{Kinks at intermediate $\beta$}

As a warm-up exercise we are going to consider the $\mathbb{CP}(1)\;$ limit \eqref{CP1_limit}. 
In the intermediate domain  $\delta m_{12} \ll \Lambda_{CP(1)} \ll \Delta m$ (or, equivalently, $1 \ll \beta \ll \beta_{AD}$), the effective \cpone model is at strong coupling, but at the same time we can use the large-$\beta$ expansion. 
The solutions of the vacuum equation \eqref{2d_equation} are given by $\sqrt{2}\sigma_\pm \approx - \Delta m / 2 \pm \Lambda_{CP(1)}$,
which yields two mirror vacua:
\begin{center}
\begin{tabular}{ c | c }
  $Vac_1$ at $\sigma = \sigma_+$ & $Vac_2$ at $\sigma = \sigma_-$ \\[2mm]
  \hline
  $X_1 \approx X_2 \approx \frac{\Lambda_{CP(1)}}{\Lambda}$ & $X_1 \approx X_2 \approx - \frac{\Lambda_{CP(1)}}{\Lambda}$ \\[2mm]
  $Y_3 \approx Y_4 \approx - \frac{\Delta m}{\Lambda}$ & $Y_3 \approx Y_4 \approx - \frac{\Delta m}{\Lambda}$
\end{tabular}
\end{center}
For both vacua the constraint \eqref{mirror_constraint} is satisfied:
\begin{equation}
	\prod_P X_P = e^{- 2 \pi \beta} \prod_K Y_K \approx \frac{\Lambda_{CP(1)}^2}{\Lambda^2} \,.
\end{equation}
There are different types of kinks interpolating between these vacua.


\begin{figure}[h!]
    \centering
    \begin{subfigure}[t]{0.4\textwidth}
        \centering
        \includegraphics[width=\textwidth]{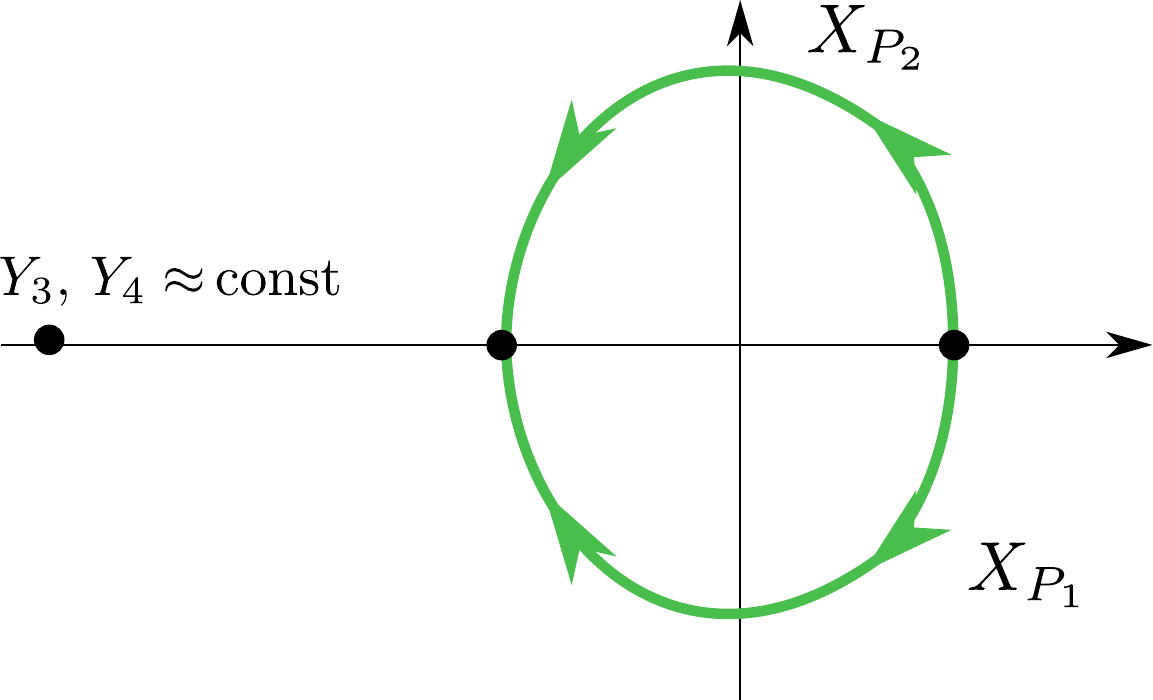}
        \caption{$P$-kinks}
        \label{fig:mirror_kinks_intermediate:P}
    \end{subfigure}%
    ~
    \begin{subfigure}[t]{0.4\textwidth}
        \centering
        \includegraphics[width=\textwidth]{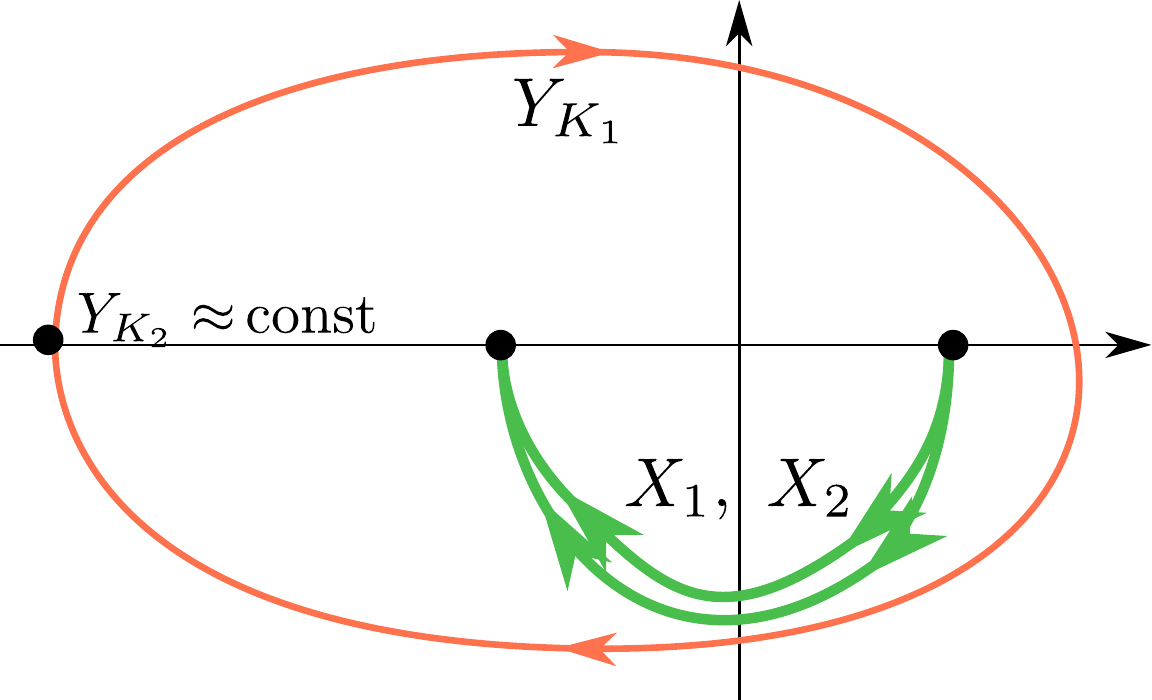}
        \caption{$K$-kinks}
        \label{fig:mirror_kinks_intermediate:K}
    \end{subfigure}%
\caption{
	Trajectories of $X_P$ and $Y_K$ in the mirror representation of a kink at intermediate $1 \ll \beta \ll \beta_{AD}$
}
\label{fig:mirror_kinks_intermediate}
\end{figure}

\paragraph{$n$-kinks} For these kinks, the two $X_P$ wind in the opposite directions, while the $Y_K$ stay intact to preserve the constraint \eqref{mirror_constraint}; see Fig.~\ref{fig:mirror_kinks_intermediate:P}. Since the arguments of $X_P$ change, the logarithms $\ln X_K$ in 
\eqref{W_mirror} acquire imaginary parts. There are two kinks of this type, depending on which flavor winds clockwise and which counter clockwise. 
From the central charge formula \eqref{mirror_kink_mass} we obtain for  these kinks
\begin{equation}
	Z_P =  \frac{2 \Lambda_{CP(1)}}{\pi} + i(m_P - \ov{m})  \,,
	\quad
	P = 1,2 \,.
\label{kink_mass_P_CP1_1}	
\end{equation}
The average mass $\ov{m}$ is defined in \eqref{mbar_mtilde}, and we used that $m_1 - m_2 = 2 (m_1 - \ov{m}) = - 2 (m_2 - \ov{m})$. 
The corresponding kink masses are  given by  the absolute values of central charges in \eqref{kink_mass_P_CP1_1}.

This formula is known in strongly coupled $\mathbb{CP}(1)\;$ and can be derived by expanding the central charge \eqref{CP1_Z} in powers of the small parameter $\delta m_{12} / \Lambda_{CP(1)}$. Namely, the  central charge \eqref{CP1_Z} reduces to the central charge of $P=1$ kink in \eqref{kink_mass_P_CP1_1} at small $\delta m_{12}$.

In the limit of equal $m_1$ and $m_2$  ($\delta m_{12}=0$) two kinks in \eqref{kink_mass_P_CP1_1} degenerate and form a doublet of the first SU(2) in the global group \eqref{globgroup}, namely
\beq
n{\rm -kinks}: \quad \left(\textbf{2},\,\textbf{1},\, 0 \right)
\label{n_kink_rep}
\eeq
The fact that kinks of the $\mathbb{CP}(N-1)\,$ model  at strong coupling form a fundamental representation of SU$(N)$ and transform as $n^P$ fields was discovered by Witten long ago \cite{W79}. Later it was confirmed by Hori and Vafa  \cite{HoVa} using the mirror representation. This is reflected in our notation of kinks in \eqref{kink_mass_P_CP1_1} as $n$-kinks. 

\paragraph{$\rho$-kinks} For these kinks, the two $X_P$ wind in one directions, while exactly one of the $Y_K$ winds double in the same direction according to \eqref{mirror_constraint}; see Fig.~\ref{fig:mirror_kinks_intermediate:K}. Then the corresponding logarithms in \eqref{mirror_kink_mass} acquire imaginary parts. There are again two kinks of this type, depending on which flavor $Y_K$ winds.
The kink central charges  are given by
\begin{equation}
	Z_K =  \frac{2 \Lambda_{CP(1)}}{\pi} + i(m_K - \ov{m})  \,,
	\quad
	K = 3,4 \,.
\label{kink_mass_K_CP1_1}
\end{equation}
These are new states, not present in $\mathbb{CP}(1)$. 
At $\beta \gg 1$ these states are much heavier than the $n$-kinks.

In the limit of equal $m_3$ and $m_4$  ($\delta m_{34}=0$) the two kinks in \eqref{kink_mass_K_CP1_1} degenerate and form a doublet of the second SU(2) in \eqref{globgroup}, namely
\beq
\rho{\rm -kinks}: \quad \left(\textbf{1},\,\textbf{2},\, 1\right).
\label{rho_kink_rep}
\eeq
These kinks behave as $\rho$ fields, see \eqref{repsnrho}. In what follows we will heavily use the fact that $n$-kinks and $\rho$-kinks transforms as
$n^P$ and $\rho^K$  fields.

Note that the BPS spectrum of \wcpt model at strong coupling is very different from that at weak coupling. First, there are no
perturbative states at strong coupling. Second, instead of the infinite tower of dyonic kinks \eqref{M2_tower} present at weak coupling
at strong coupling we have just four kinks which belong to representations  \eqref{n_kink_rep} and \eqref{rho_kink_rep}
of the global group \eqref{globgroup}. Note also that  global charges of kinks in the perturbative tower \eqref{M2_tower}
associated with the single mass difference $(m_1 -m_2)$. In contrast the kink global charges at  strong coupling are associated with all masses $m_A$ present in the model.
We study CMS where the transformations of the BPS spectra occurs in Sec.~\ref{sec:CMS}.

The above  results can be directly generalized to the dual domain of negative $\Re\beta$.  
When $\beta$ is in the intermediate domain between $- |\beta_{AD}|$ and $-1$, the  $n^P$  fields are heavy and decouple, and we are again left with a $\mathbb{CP}(1)\;$ model, only this time comprised of the  $\rho^K$ fields and a new strong coupling scale
\begin{equation}
	\Lambda_{\wt{CP}(1)} = \Delta m \, e^{+ \pi \beta} \,.
\end{equation}
Roles of $n$-kinks and $\rho$-kinks are reversed. Their central charges  are given by
\begin{equation}
	Z_P =  \frac{2 \Lambda_{\wt{CP}(1)}}{\pi} + i(m_P - \wt{m}) 
	\ , \quad
	Z_K =  \frac{2 \Lambda_{\wt{CP}(1)}}{\pi} + i(m_K - \wt{m}) 
	\,.
\label{kink_mass_wtCP1_1}	
\end{equation}
with $\wt{m}$ defined in \eqref{mbar_mtilde}. As we can see, now the $Z_K$-kinks are light. 
Note that these results match with the $S$-duality transformation \eqref{2d_S-duality}.

Finally, we note that apart from the kinks just described, there can be kinks described by $X_P,\ Y_K$ fields winding in the opposite direction. Say, for $\rho$-kinks on Fig.~\ref{fig:mirror_kinks_intermediate:K} the $X_P$ may wind in the upper half plane, with $Y_K$ winding counter clockwise. These kinks turn out to be $n=+1$ states from  the strong coupling tower of higher winding  states discussed in  the next subsection.

\subsection{Kinks near the origin $\beta = 0$}

\begin{figure}[h]
    \centering
    \begin{subfigure}[t]{0.4\textwidth}
        \centering
        \includegraphics[width=\textwidth]{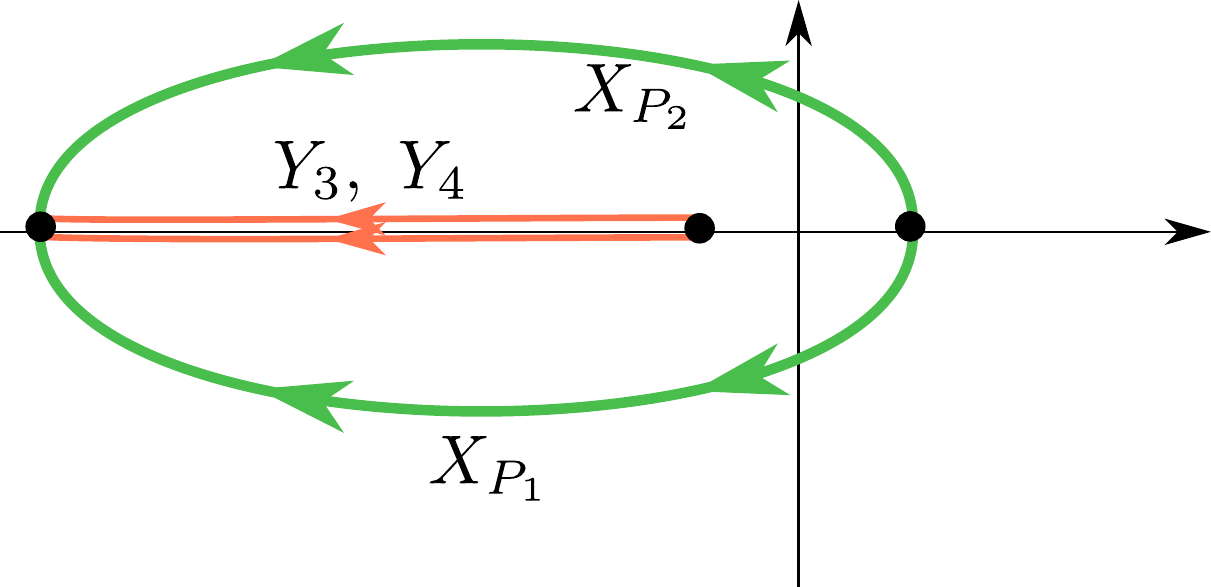}
        \caption{$P$-kinks}
        \label{fig:mirror_kinks_origin:P}
    \end{subfigure}%
    ~
    \begin{subfigure}[t]{0.4\textwidth}
        \centering
        \includegraphics[width=\textwidth]{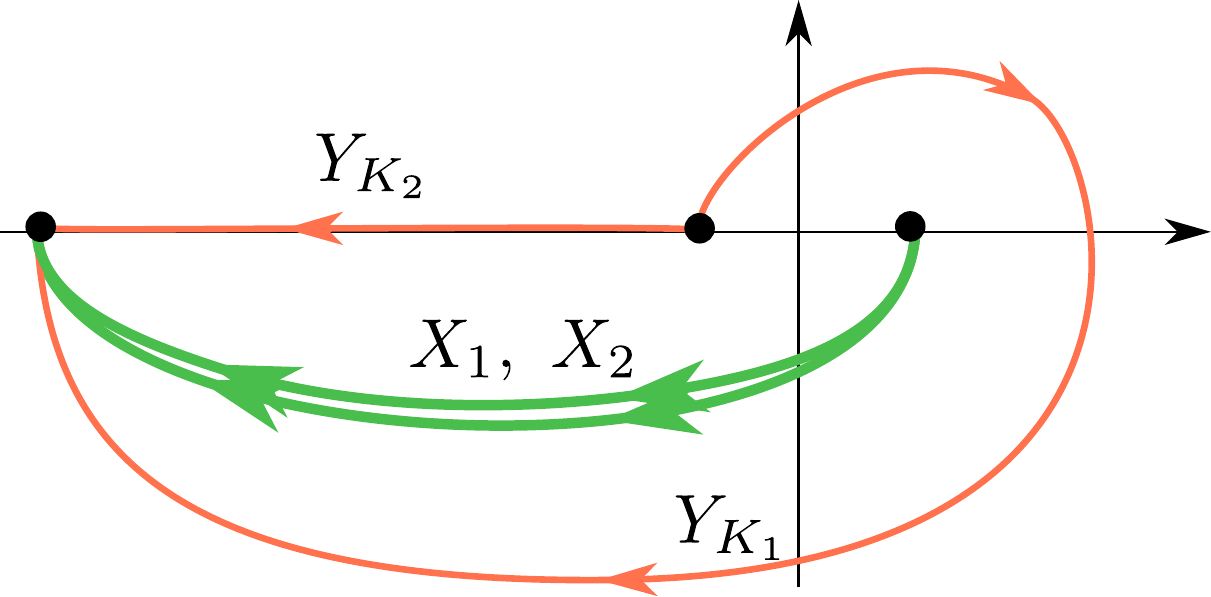}
        \caption{$K$-kinks}
        \label{fig:mirror_kinks_origin:K}
    \end{subfigure}%
\caption{
	Trajectories of $X_P$ and $Y_K$ in the mirror representation of a kink at $\beta \to 0$.
}
\label{fig:mirror_kinks_origin}
\end{figure} 

Now consider the limit $\beta \to 0$. 
In the vicinity of the origin the last condition in \eqref{CP1_limit} is badly broken, and all of the \wcpt fields $n^P$, 
$\rho^K$ \eqref{wcp22} play an important role.

In this limit we can use the small-$\beta$ expansion. We have $e^{- 2 \pi \beta} \approx 1 - 2 \pi \beta$, and the $\sigma$-vacua \eqref{roots_symmetric} are approximately
\begin{equation}
	\sqrt{2} \sigma_+ \approx \frac{\delta m_{12}^2 - \delta m_{34}^2}{8 \Delta m ^2} 
	\,, \quad 
	\sqrt{2} \sigma_- \approx - \frac{\Delta m}{\pi \beta} \,.
\label{mirror_roots_approx_beta=0}
\end{equation}
Without loss of generality we can consider the limit when $\sigma_+ \approx 0$. Then, the two mirror vacua are given by
\begin{center}
\begin{tabular}{ c | c }
  $Vac_1$ at $\sigma = \sigma_+$ & $Vac_2$ at $\sigma = \sigma_-$ \\
  \hline
  $X_P \approx m_P  / \Lambda$ & $X_1 \approx X_2 \approx - \frac{\Delta m}{\pi \beta \Lambda}$ \\
  $Y_K\approx m_K  / \Lambda$ & $Y_3 \approx Y_4 \approx - \frac{\Delta m}{\pi \beta \Lambda}$
\end{tabular}
\end{center}

Again, there are two types of kinks.
$n$-kinks are obtained when, say, $X_1$ picks up the phase $+i\pi$, $X_2$ picks up $-i\pi$, while the phases of $Y_K$ remain intact; see Fig.~\ref{fig:mirror_kinks_origin:P}. There is also a kink for which the roles of $X_1$ and $X_2$ are reversed. We have the total of two kinks with central charges
\begin{equation}
	Z_P =  \frac{m_1 + m_2 - m_3 - m_4}{2 \pi} \ln \frac{2}{\pi \beta} + i (m_P - \ov{m})
\label{kink_mass_P_beta0}
\end{equation}
where $\ov{m}$ is defined in \eqref{mbar_mtilde}.

Similarly, the $\rho$-kinks are obtained when two $X_P$ wind with the same phase, while exactly one of the $Y_K$ winds twice as much in accord with \eqref{mirror_constraint}; see Fig.~\ref{fig:mirror_kinks_origin:K}. The central charges  of these kinks are given by
\begin{equation}
	Z_K =   \frac{m_1 + m_2 - m_3 - m_4}{2 \pi} \ln \frac{2}{\pi \beta} + i (m_K - \ov{m})  \,.
\label{kink_mass_K_beta0}
\end{equation}
We immediately observe that the kink masses are singular at the self-dual point $\beta=0$. They are very heavy in the vicinity of this point.

To get the kink spectrum at ${\rm Re}\beta <0$ we can analytically continue  \eqref{kink_mass_P_beta0} and
\eqref{kink_mass_K_beta0} to $\beta \to \tilde{\beta}=-\beta$. The log terms in \eqref{kink_mass_P_beta0}, 
\eqref{kink_mass_K_beta0} give $i(m_1 + m_2 - m_3 - m_4)/2$ which converts $\bar{m}$ into $\tilde{m}$.
Note, that this matches with the $S$-duality transformation \eqref{2d_S-duality}.

Now observe that the central charges of $n$ and $\rho$-kinks \eqref{kink_mass_P_beta0} and \eqref{kink_mass_K_beta0} have a branching point at $\beta=0$. This is a new feature absent in asymptotically free versions of \wcp models. What is the meaning of this branching point? Below in this section we will ague that the self-consistency
of the BPS spectrum requires the presence of a new tower of  higher winding states in our conformal \wcpt model.
This tower is present  only at strong coupling and decays as we move to large $\beta$. This can be seen as follows.

Consider changing the coupling constant  $\beta$ along some trajectory in the complex plane. This trajectory may stretch from the weak coupling region $\beta \gg \beta_{AD}$ through the strong coupling domain into the dual weak coupling region $\beta \ll -\beta_{AD}$. This trajectory may also encircle an AD point and go through a cut on a different sheet. The charges of various BPS states change, but there are CMS starting at the AD points, and the BPS spectrum as a whole stays intact. The would-be 
\textquote{extra} states decay on CMS \cite{DoHoTo}.

However, this trajectory may also go full circle around the singularity $\beta=0$. It can also encircle this point several times. There are no CMS starting at $\beta=0$ and extending outwards. What we end up with is another set of BPS states. From the expressions for the kink central charges \eqref{kink_mass_P_beta0}, \eqref{kink_mass_K_beta0} we see that if we go around the origin $n$ times, then the central charge of the BPS kinks becomes
\begin{equation}
\begin{aligned}
	Z_A^{[n]} &=  \frac{m_1 + m_2 - m_3 - m_4}{2 \pi} \ln \frac{2}{\pi \beta} + i (m_A - \ov{m}) + i \, n \cdot (m_1 + m_2 - m_3 - m_4) \,, \\
	&\frac{\pi}{2} \leqslant \arg\beta < - \frac{3 \pi}{2} \,.
\end{aligned}
\label{Z_higher_windings}
\end{equation}
Here the argument of $\beta$ is constrained so as to account for the cut, see Fig.~\ref{fig:CMS_right_13}.
Does it mean that the full BPS spectrum changes as we go to other sheets?

The way to resolve this issue is to assume that that in fact {\em all} of the states \eqref{Z_higher_windings} are already present at strong coupling on the first sheet. When we wind circles around the origin, this tower of states $Z_A^{[n]}$ simply shifts in the index $n$. Since this index runs over all integers and the number of states in the tower is infinite, the whole BPS spectrum is in fact $2\pi$-periodic with respect to $\arg\beta$.

The new tower \eqref{Z_higher_windings} is present only at strong coupling. At weak coupling it decays. We study associated CMS
and decay processes in Appendix~\ref{sec:higher_winding}.

%
%

\section{CMS}
\label{sec:CMS}

In this section we will present the curves of marginal stability (CMS) for various decays of BPS states.

As was stated above, in \wcpt theory under consideration  the coupling $\beta$ does not run.
We want to understand transformations of the BPS spectrum at different values of $\beta$, particularly weak vs. strong coupling regions as well as at  $\Re\beta < 0$. In order to better capture the relevant effects, we are going to investigate more closely how the particle spectrum depends on $\beta$ while holding the masses\footnote{Or, rather, their ratios since the CMS positions on the $\beta$ plane can depend only on dimensionless parameters, and there is no dynamical strong coupling scale $\Lambda$ in \wcpt.} 
$m_A$ fixed. 
To this end we will study curves of marginal stability (CMS) on the complex $\beta$ plane.
Since the $\theta_{2d}$ angle is $2\pi$ periodic, the whole picture of spectra will be periodic as well.

In Sections \ref{sec:weak} and \ref{sec:mirror} we saw that the strong and weak coupling spectra are different. At weak coupling $\beta \gg \beta_{AD}$, we observed  the dyonic tower \eqref{M2_tower} as well as ``perturbative" states with the central charge $i(m_1 - m_2)$. They are not present at strong coupling and must decay on a CMS separating the strong and  weak coupling regions. We will refer to these CMS as the {\em }primary curves.

Moreover, at strong coupling we have $\rho$-kinks not present at weak coupling.
Correspondingly, CMS must exist on which these states will decay. We will call these the {\em secondary curves}.

Finally, we saw that at weak coupling there are the so-called bifundamentals -- the perturbative  states with masses $|m_P - m_K|,\ P=1,2,\ K=3,4$. These states do not decay even at strong coupling. They are present everywhere on the $\beta$ plane. To see that  this is the case
suffice it to note that in the massless limit $m_A \to 0$ 4D SQCD has a Higgs branch formed by bifundamental quarks. This Higgs branch is protected by supersymmetry and present at all couplings. Through the 2D-4D correspondence we conclude that 2D bifundamentals are also present at all $\beta$.

Below in this section we study the primary CMS while the are secondary CMS discussed in Appendix~\ref{sec:secondary_CMS}.

\begin{figure}[h!]
	\centering
	\includegraphics[width=0.7\linewidth]{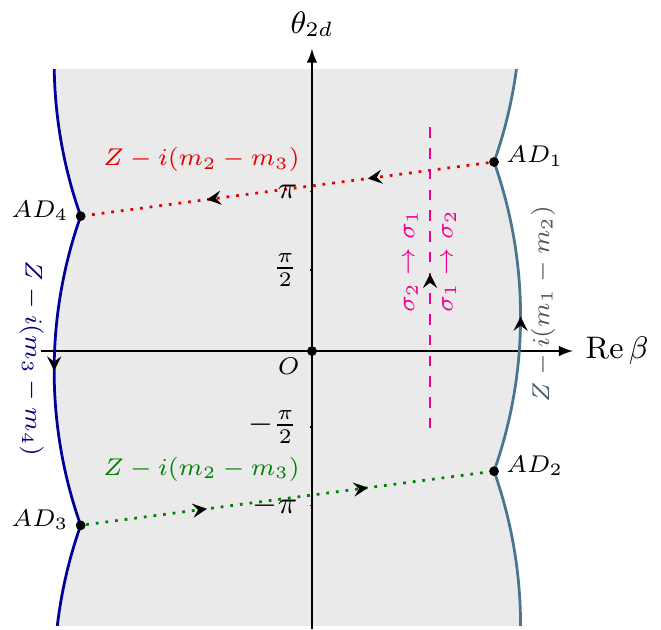}
	\caption{
		Primary CMS (solid lines on the left and on the right, schematically) and central charge shifts (see Appendix~\ref{sec:Z_windings}).
		$AD_A$ are the Argyres-Douglas points where the corresponding central charge $Z_A$ vanishes.
		The  masses $m_A$ are generic.
		Grey region is the strong-coupling domain.
		 When going from one AD point to the next shown in this  figure one observes phase shifts.
		There is also a $\mathbb{Z}_2$ 
			transformation (exchanging the $\sigma$ roots) when shifting $\theta_{2d} \to \theta_{2d} + 2\pi$.
		Apart from that, the picture is $2 \pi$ periodic with respect to $\theta_{2d}$.
		$AD_2 = AD_1 - 2 \pi i$, $AD_3 = AD_4 - 2 \pi i$.  
	}
\label{fig:CMS_type1}
\end{figure}

\subsection{Primary curves in the \boldmath{$\beta$} plane}

As was discussed above, when we pass from large $\beta \gg \beta_{AD}$ to strong coupling $\beta \sim 1$, the perturbative states with the central charge $i(m_1 - m_2)$ decay on CMS producing (dyonic) kink-antikink pairs. We can write the decay processes schematically as%
\footnote{Here and further on we use the notation with square brackets $[Z_A]$, $[i (m_1 - m_2)]$ to represent particles with the corresponding central charges. The central charge of an antiparticle equals negative of that of the particle.}
\begin{equation}
	\underbrace{[ i (m_1 - m_2) ]}_{\text{elementary quantum}} \to 
		\underbrace{[Z_1]}_{\text{dyon}}
		+ \underbrace{[- Z_2]}_{\text{antidyon}} \,.
\label{right_decay_W}		
\end{equation}
On the CMS, the central charges of the decaying particles must have the same argument, i.e. they must be collinear vectors in the complex plane. From this we can derive the equation for the CMS,
\begin{equation}
	\Im \left( \frac{ Z_P }{i (m_1 - m_2)} \right) = 0 
	\Leftrightarrow
	\Re \left( \frac{ Z_P }{m_1 - m_2} \right) = 0
	\,, \quad
	P = 1, \, 2 \,.
\label{CMS_right_curve}		
\end{equation}
The same decay curve describes the decay of the dyonic tower  \eqref{M2_tower} into the strong coupling states.

This curve separates the weak coupling region $\beta \gg \beta_{AD}$ from the strong coupling region. It passes through the AD points \eqref{AD_roots_simplecase_beta1_approx} with ${\rm Re}\,\beta >0$ where the mass of one of the $n$-kinks $[Z_P]$   vanishes. We denote these AD points AD$_P$, $${\rm AD}_2 ={\rm AD}_1 - 2\pi i\,,$$ see Appendix~\ref{sec:Z_windings} for a detailed discussion. Of course, the corresponding CMS is $2\pi$ periodic in $\theta_{2d}$. 

We solve \eqref{CMS_right_curve} numerically. The result is presented by 
 the r.h.s. curve on Fig.~\ref{fig:CMS_type1}. Note that $[Z_1]$ and  $[Z_2]$ kinks 
\eqref{kink_mass_P_CP1_1} present at strong coupling survive at the weak coupling region at positive $\beta\gg \beta_{AD}$.
In this region they belong to the tower  \eqref{M2_tower} with $n=0$ and $n=-1$ respectively. This is a well-known behavior: the states which can become massless at some points are present both at weak and strong coupling \cite{SW1,Ferrari:1996sv}.  

Note that the CMS curve for \cpone model in the complex plane $\delta m_{12}$ is well-known \cite{ShVanZwi}. 
It is a closed curve around the origin which passes through the AD points. Our  curve in Fig.~\ref{fig:CMS_type1} (more exactly, its right branch at  ${\rm Re}\,\beta >0$) is a translation of the  curve in  \cite{ShVanZwi} into the 
$\beta$ plane. In the $\beta$ plane the curve is not closed. It is periodic in $\theta_{2d}$.

Analogously, when $\Re\beta$ is large but negative (l.h.s. on Fig.~\ref{fig:CMS_type1}), there are perturbative states whose central charge is $i(m_3-m_4)$ and a corresponding dyonic tower. Their decay curves satisfy
\begin{equation}
	\Re \left( \frac{ Z_K }{m_3 - m_4} \right) = 0
	\,, \quad
	K = 3, \, 4 \,.
\label{CMS_left_curve}		
\end{equation}
This CMS separates the dual weak coupling region $\beta \ll - \beta_{AD}$ from the strong coupling region. 
On Fig.~\ref{fig:CMS_type1} it is drawn on the left side.

We see two weak coupling regions in the complex plane of $\beta$ . They are separated by a strong coupling region which resembles a band stretched along the $\theta_{2d}$ direction. This is illustrated on Fig.~\ref{fig:CMS_type1}.

%
%

\section{Instead-of-confinement phase }
\label{sec:instead_of_conf}

In this section we use 2D-4D correspondence to confirm the instead-of-confinement phase in the bulk 4D SQCD at strong coupling. This phase  was discovered earlier in asymptotically free versions of SQCD \cite{SYdual}, see \cite{SYdualrev} for a review.

To this end we first consider our world-sheet \wcpt model on the non-Abelian string. In the previous sections  we have learned that 
the BPS spectrum of states is very different at weak and strong coupling. In particular, the perturbative states with
mass $|m_1-m_2|$ decay into say, $[Z_1]$ kink and $[-Z_2]$ antikink on the CMS on the the r.h.s. in Fig.~\ref{fig:CMS_type1} when we 
pass from the weak coupling region into the strong coupling one. At strong coupling these perturbative states do not exist.

The 2D-4D correspondence tells us that a similar process occurs on the Coulomb branch (at $\xi =0$) in 4D SQCD when we pass from the weak coupling region to the strong coupling one. The 2D perturbative states with mass $|m_1-m_2|$ correspond in the bulk theory (4D SQCD) to
BPS off-diagonal) quarks $q^{kP}$, $P=1,2$, and  gluons. They do not exist at strong coupling. They decay into monopole and 
anti-monopole pair\,\footnote{We call all 4D states with non-zero magnetic charge monopoles although they can be dyons  carrying
also electric and global charges \cite{SW2}.}. 

Moreover, since the $n$-kinks of the 2D theory form doublets with respect to
the first SU(2) factor of the global group \eqref{globgroup}, see \eqref{n_kink_rep} and \eqref{rho_kink_rep}, the  monopoles and anti-monopoles formed as a result of the quark/gluon decay also transform as doublets and anti-doublets of the first SU(2) factor of the global group.

As we turn on $\xi$ at  weak coupling the 4D theory goes into the Higgs phase. Quarks $q^{kP}$, $P=1,2$, get screened by the condensate \eqref{qvev}. They combine with massive gluons to form a long non-BPS \ntwo multiplets with mass $g\sqrt{\xi}$,
see the review \cite{SYrev} for details. Moreover, at non-vanishing values of  $\xi$
the monopoles become confined by non-Abelian strings.

 Now, if we move to the strong coupling domain, the monopole and anti-monopole
created as a result of the quark/gluon decay cannot move apart. They are attached to two confining strings and form 
a monopole-antimonopole stringy meson shown in Fig.~\ref{monmb}a. Of course this meson is a non-BPS state. Its mass is of the
order of $\sqrt{\xi}$. Note, that this meson is formed also in the massless limit $m_A\to 0$. The mass scale in 4D SQCD is set by the FI 
parameter $\xi$. 

Thus we see that the screened quarks and gluons present in 4D SQCD in the Higgs phase at weak coupling do not survive when we move to strong coupling. They evolve into monopole-antimonopole stringy mesons. We call the phase which emerges at the strong coupling  {\em instead-of-confinement} phase \cite{SYdualrev}.

This phase is  an alternative to the ordinary confinement phase in QCD. The role of the constituent quarks in this phase is played by confined monopoles. Moreover, since monopoles and antimonopoles  transform as doublets and antidoublets of the first SU(2) factor of the global group \eqref{globgroup_d=4} the stringy mesons  appear in the singlet  or adjoint representations 
of the first SU(2) subgroup. This is similar to what happens in QCD: quark-antiquark mesons form the singlet or adjoint representation of the flavor group.

The same instead-of-confinement mechanism  works  if we start at large negative $\beta $ and pass through l.h.s CMS into strong coupling.
The monopole-antimonopole stringy mesons formed on this CMS appear in the singlet  or adjoint representations 
of the second SU(2) subgroup of the global group.
The strong coupling region between the r.h.s and l.h.s. CMS in Fig.~\ref{fig:CMS_type1} in 2D theory corresponds to 
the strong coupling domain around the large semicircle in Fig.~\ref{fig:coupling_traj} in terms of the complexified 4D
coupling $\tau$, see \eqref{tau_def} in Sec.~\ref{sec:2D_4D}. This is the region of instead-of-confinement phase in 4D
SQCD.

%
%

\section{Stringy baryon from field theory \label{sec:baryon_2d}}


In this section we show that the presence of the baryonic state \eqref{brep} found as a  massless string state of the critical string theory on the non-Abelian vortex in our 4D \ntwo SQCD can be confirmed  using purely field-theoretical methods.
Let us start with the world-sheet \wcpt model on the string at strong coupling near the origin in the $\beta$ plane.
The baryonic charge $Q_B=2$ and the absence of the Cartan charges with respect to both SU(2) factors of the global group 
\eqref{globgroup} suggests  that this state can be formed as a BPS bound state of two different $n$-kinks and two different $\rho$-kinks arranged  on the infinite straight  string  in the following order
\beq
[Z_P]|_{1\to 2} + [Z_K]|_{2\to 1} + [Z_{P'}]|_{1\to 2} + [Z_{K'}]|_{2\to 1}, \qquad P\neq P', \quad K\neq K'\,,
\label{4_kinks}
\eeq
where the subscript $|_{1\to 2}$ ($|_{2\to 1}$) denotes the   kink  interpolating from vacuum 1  to vacuum 2 (vacuum 2  to vacuum 1).
The central charges of the second and last kinks come with the minus sign, see \eqref{mirror_kink_mass}, and the net central charge 
of the bound state \eqref{4_kinks} is 
\beq
Z_b = i(m_1 + m_2 -m_3 -m_4),
\eeq
see \eqref{kink_mass_P_beta0} and \eqref{kink_mass_K_beta0}.
Note that this state cannot have a net topological charge. 
The 2D  topological charge translates into  4D  magnetic charge of a monopole.
Clearly, the baryon (or any other hadron) cannot have 
color-magnetic charge because  magnetic charges are confined in 4D SQCD.

The 4-kink composite state \eqref{4_kinks} transforms under the global group \eqref{globgroup} as 
\begin{equation}
	n^P \rho^K n^{P'} \rho^{K'} = w^{PK} w^{P'K'},
\label{baryon_candidate}		
\end{equation}
where we use the gauge invariant mesonic variables \eqref{w}.
  It is clear that \eqref{baryon_candidate} is symmetric with respect to indices $P,P'$ and $K,K'$. Thus, this state is in 
	the triplet representation  ({\bf 3}, {\bf 3}, 2) of the global group. This is not what we need.
	
The singlet representation ({\bf 1}, {\bf 1}, 2) \eqref{brep} we are looking for would correspond to $\det(w)$.
But it is zero, see \eqref{coni}! 

However, recall that it is zero only in \wcpt model formulated in terms
of  $n$'s and $\rho$'s. 
Let us take the massless limit $m_A\to 0$ and go to the point $\beta=0$.
Our world-sheet \wcpt  theory on the conifold allows a marginal deformation of the conifold complex structure
 at $\beta =0$ \cite{NVafa,Candel}, namely 
\beq
\det (w) = b
\label{deformedconi1},
\eeq
where $b$ is a complex parameter, see \eqref{deformedconi} in Sec.~\ref{conifold}.
  This deformation preserves Ricci-flatness which ensures that 2D world-sheet  theory is still conformal and has no
dynamical	$\Lambda$ scale, so the   baryonic state $\det (w)$ which emerges in the deformed theory is massless. 
	
Next we use the 2D-4D correspondence that ensures that at $\beta=0$ and non-zero $b$ there is a similar massless baryonic BPS state in
 4D SQCD formed by  four monopoles.  At non-zero values of $\xi$, the monopoles are confined and  this baryon is represented by  a
necklace configuration formed by four monopoles connected by confining strings, see Fig.~\ref{monmb}b. At non-zero $\xi$ this state becomes a well-defined localized state in 4D SQCD. Its size is determined by $1/\sqrt{\xi}$. Note, that this 
baryon is still a short massless BPS hypermultiplet at nonvanishing $\xi$ because there is no other massless BPS state with the same quantum numbers to combine with to form a long multiplet \footnote{This is similar to what happens with the bifundamental quarks which
remain massless BPS states as we switch on a non-zero $\xi$.}.

Now we can address the question: what is the origin of the marginal deformation parameter $b$ in 4D SQCD? As was already mentioned in  
Sec.~\ref{sec:introduction}, it can be a marginal coupling constant which respects \ntwo supersymmetry or a VEV of a dynamical state. The coupling constant
$\beta$ is associated with the deformation of the K\"ahler class of the conifold rather then its complex structure. Moreover, note
that the deformation parameter $b$ cannot be a coupling associated with gauging of any symmetry of the global group 
\eqref{globgroup_d=4}
because it has non-zero $Q_B$.
This leads us to the conclusion that $b$ is a VEV of a dynamical state, namely the VEV of the massless stringy four-monopole baryon discussed above.

The baryon $b$ exists only at the origin $\beta = 0$. As we move away from $\beta=0$, it must decay on a point-like degenerate CMS which tightly wraps the origin. It decays into two massless bifundamental quarks which belong to the representation ({\bf 2}, {\bf 2}, 1) of the global group.

Thus we confirm that a new non-perturbative Higgs branch of real dimension ${\rm dim}\,{\cal H}= 4$ opens up in our 4D SQCD at the point $\beta=0$ (up to $2\pi$ periodicity of the $\theta_{2d}$ angle) in the massless limit.
Most likely the perturbative Higgs branch \eqref{dimH} formed by bifundamental quarks is  lifted  at $b\neq 0$.
The point $\beta=b=0$ is a  phase transition point, a singularity where two Higgs branches meet. This issue needs future clarification.

%
%

\section{Detailing the 2D-4D correspondence \label{sec:2D_4D}}

As was stated above, the sigma model \eqref{wcp22} is an effective world-sheet theory on the semilocal non-Abelian string in four-dimensional \ntwo SQCD. Generally speaking, if we consider the bulk theory with the gauge group U($N$) and $N < N_f \leqslant 2N$ flavors of quarks, then the world-sheet theory is the weighted sigma model \WCP. In this paper we focus on the case $N_f = 2N = 4$.

The mass parameters $m_A$ of the world-sheet theory \eqref{wcp22} are the same as quark masses in the bulk 4D SQCD. The two-dimensional coupling $\beta$ \eqref{beta_complexified} is also related to the four-dimensional complexified coupling constant $\tau_\text{SW}$ which is defined as
\begin{equation}
	\tau_\text{SW} = i \, \frac{8\pi}{g^2}  + \frac{\theta_{4d}}{\pi} \,.
\label{tau_def}	
\end{equation}
Here $\theta_{4d}$ is the four-dimensional $\theta$ angle. 
We will start this section from derivation of the corresponding relation.

\subsection{Relation between the couplings}
\label{rbc}

In the weak coupling limit, the known classical-level relation between the couplings of the bulk and world-sheet theories
is \cite{SYmon,ABEKY}
\begin{equation}
	\Re\beta \approx \frac{4 \pi}{g^2} \,.
\end{equation}
But what is the exact formula?

To establish a relation applicable at the quantum level, we are going to use the 2D-4D correspondence -- the coincidence 
of the BPS spectra of monopoles in 4D SQCD and  kinks in 2D world-sheet $\mathbb{WCP}(N,\tN)$ model, see Sec.~\ref{2-4}. As was already noted  the key technical reason behind  this
coincidence is that the VEVs of $\sigma$ given by the exact twisted superpotential  coincide
with the double roots of the Seiberg-Witten curve \cite{SW2} in the quark vacuum of
the 4D SQCD \cite{Dorey,DoHoTo}. Below we use this coincidence  to derive the exact relation between 4D coupling $\tau_{SW}$ and 2D coupling $\beta$ in the theory at hand, $N_f = 2N = 4$, where both couplings do not run.

 Mathematically, this can be formulated as follows. Consider the Seiberg-Witten curve of the bulk SQCD.
The Seiberg-Witten (SW) curve for the SU($N$)  gauge theory with $N_f = 2 N$ flavors was derived in \cite{ArgPlessShapiro,APS}. 
It has the form
\begin{equation}
	y^2 = \prod_{a=1}^{N}(x-\phi_a)^2 + h(h+2) \prod_{i=1}^{N_f} (x + h m_S + m_i)
	\,, \quad
	N_f = 2 N \,.
\label{4d_curve_sun}	
\end{equation}
Here, $$h \equiv h(\tau_\text{SW})$$  is a modular function \eqref{h_def_APS},
see Appendix~\ref{sec:modular_101}.
Moreover, $\tau_\text{SW}$ is defined in \eqref{tau_def}). The parameter $m_S$ in (\ref{4d_curve_sun}) is the average mass, 
\begin{equation}
	m_S = \frac{1}{N_f} \sum_{i=1}^{N_f} m_i \,.
\label{m_avg}	
\end{equation}
The combination $h(h+2)$ is invariant under $S$ and $T$ duality transformations.

In fact, we are interested in the case when the gauge group is actually 
\beq
\mbox{U$(N)=$SU($N$) $\times$ U(1)}\,.
\eeq
 Therefore we can make a shift
$x\to (x + hm_S)$, $\phi_a\to (\phi_a+hm_S)$ and get rid of $hm_S$. Note, that  in the U($N$) theory -- in contrast to the SU($N$) case --
the $\sum _a\phi_a$ does not have to vanish.
The SW curve \eqref{4d_curve_sun} then becomes
\begin{equation}
	y^2 = \prod_{a=1}^{N}(x-\phi_a)^2 + h(h+2) \prod_{i=1}^{N_f} (x + m_i)
	\,, \quad
	N_f = 2 N \,.
\label{4d_curve}	
\end{equation}
Our quark vacuum is  a singular point on the Coulomb branch where all the Seiberg-Witten roots are double roots, so
the diagonal quarks $q^{kP}$ with $k=P$ are massless. Upon switching on a nonvanishing $\xi$  this singularity transforms into an isolated vacuum  where the diagonal quarks develop VEVs \eqref{qvev}.
 
To guarantee the coincidence of the BPS spectra,
we require that the double roots of the four-dimensional Seiberg-Witten curve \eqref{4d_curve} coincide with the solutions of the two-dimensional vacuum equation \eqref{2d_equation}. In asymptotically free versions of the theory the SW curve is simply  the square of the vacuum equation of the two dimensional theory \cite{Dorey}. This ensures the coincidence of roots.
We use the same idea for the conformal case at hand.

Consider the square of \eqref{2d_equation} in the following form:
\begin{equation}
	\wt{y}^2 = \left[ 
			\prod_{P=1}^{N}\left(\sqrt{2} \sigma + m_P \right) 
			- e^{- 2 \pi \beta} \prod_{K = N + 1}^{2 N} \left(\sqrt{2} \sigma + m_K \right)
		\right]^2		\,.
\label{2d_to_4d_1}		
\end{equation}
We want to make a connection with the SW curve \eqref{4d_curve}. Equation \eqref{2d_to_4d_1} can be rewritten as
\begin{equation}
	\wt{y}^2 = \left[ 
			\prod_{P=1}^{N}\left(\sqrt{2} \sigma + m_P \right) 
			+ e^{- 2 \pi \beta} \prod_{K = N + 1}^{2 N} \left(\sqrt{2} \sigma + m_K \right)
		\right]^2	
		- 4 \, e^{- 2 \pi \beta} \, \prod_{A = 1}^{2 N} \left(\sqrt{2} \sigma + m_A \right)
		\,.
\label{2d_to_4d_2}		
\end{equation}
Let us compare this to the four-dimensional curve \eqref{4d_curve}. We immediately identify
\begin{equation}
\begin{aligned}
	x &= \sqrt{2} \sigma \,, \\[2mm]
	h (h + 2) &= - \frac{4 e^{- 2 \pi \beta}}{(1 + e^{- 2 \pi \beta})^2}  \,, \\[2mm]
	y^2 &=  \frac{ \wt{y}^2 }{ (1 + e^{- 2 \pi \beta})^2 }  \,.
\end{aligned}
\label{2d_4d_map_1}	
\end{equation}
We can also find the Coulomb branch parameters,
\begin{equation}
	\phi_{1,2} = - \frac{\Delta m}{2} \, \frac{1 - e^{- 2 \pi \beta}}{1 + e^{- 2 \pi \beta}} 
		~\pm~ \sqrt{ \frac{(\delta m_{12})^2 + e^{- 2 \pi \beta} \, (\delta m_{34})^2}{4 (1 + e^{- 2 \pi \beta}) } - \Delta m^2 \, \frac{e^{- 2 \pi \beta}}{(1 + e^{- 2 \pi \beta})^2}}\,,
\end{equation}
where the mass notation is according to \eqref{mass_parametrization}.
Note that one of these Coulomb parameters diverges at $\beta = i k / 2$, $k \in \mathbb{Z}$ (cf. our discussion in Appendix~\ref{sec:dual_couplings}).

The second  relation in \eqref{2d_4d_map_1} can be viewed as a quadratic equation with respect to $e^{- 2 \pi \beta}$.
Solving it, we obtain two solutions
\begin{equation}
\begin{aligned}
	e^{- 2 \pi \beta_1} &= \lambda(\tau_\text{SW} + 1)  \,,  \\
	e^{- 2 \pi \beta_2} &= \frac{1}{\lambda(\tau_\text{SW} + 1)}  \,,
\end{aligned}
\label{2d_4d_coupling_two_solutions}	
\end{equation}
where we used \eqref{h_combination} and \eqref{lambda_trans_halfshift}. See Sec.~\ref{klambdaf} for the definition of the $\lambda$ functions.
These two solutions are interchanged by the $S$ duality transformation, see \eqref{lambda_p1_S}. 

In the weak coupling limit $\Im \tau_\text{SW} \gg 1$ the $\lambda$ functions in \eqref{2d_4d_coupling_two_solutions} can be expanded according to  \eqref{lambda_function}. For the first option in \eqref{2d_4d_coupling_two_solutions} we have
\begin{equation}
	e^{- 2 \pi \beta} \approx 16 e^{\pi i (\tau_\text{SW} + 1)} \,.
\end{equation}
Recalling the definitions of the complexified couplings \eqref{beta_complexified} and \eqref{tau_def}, we can write down the weak coupling relation as
follows:
\begin{equation}
\begin{aligned}
	r &\approx \frac{4 \pi}{g^2} - \frac{2\ln(2)}{\pi}	\,, \\[2mm]
	\theta_{2d} &\approx - \theta_{4d} - \pi   \,,
\end{aligned}
\label{beta_of_tau_weak_coupling}
\end{equation}
cf. Eq. (\ref{beta_complexified}).
This is compatible with the known quasiclassical results.
From this analysis we see that out of two options \eqref{2d_4d_coupling_two_solutions}, the first one gives a correct weak coupling limit.
Thus we can write down our final formula for the relation between the world-sheet and bulk couplings \footnote{This result corrects the relation claimed previously in \cite{SYlittles} without derivation.   It should be compared with the result 
\cite{Karasik} obtained in 2017 by Gerchkovitz and Karasik. The latter is not quite identical to (\ref{2d_4d_coupling_my}). See the explanation below Eq. (\ref{2d_4d_coupling_my_noshift}).},
\begin{equation}
	e^{- 2 \pi \beta} = \lambda(\tau_\text{SW} + 1) \,.
\label{2d_4d_coupling_my}	
\end{equation}
To visualize this relation between 4D and 2D couplings see Fig.~\ref{fig:coupling_traj} and Fig.~\ref{fig:beta_traj}.

Note, that different possible forms of the SW curve can lead to different relations between 4D and 2D couplings. For example, 
in \cite{ArgPlessShapiro} the authors claimed that the shift $\tau_\text{SW} \to \tau_\text{SW} + 1$ is basically a change of the origin of the $\theta$ angle by $\pi$, so supposedly it does not change physics, but only changes the form of the SW curve.
The curve \eqref{4d_curve} corresponds to the choice
\begin{equation*}
	g = \frac{\theta_2^4 + \theta_1^4}{\theta_2^4 - \theta_1^4} ,
	\quad
	h (h + 2) = - (1 - g^2) \,.
\end{equation*}
for the function $g$ from \cite{ArgPlessShapiro}. One could have also chosen this function differently, 
\begin{equation*}
	g = \frac{\theta_3^4 - \theta_1^4}{\theta_3^4 + \theta_1^4} \,,
\end{equation*}
which would lead to 
\begin{equation}
	e^{- 2 \pi \beta} = \lambda(\tau_\text{SW})
\label{2d_4d_coupling_my_noshift}	
\end{equation}
instead of \eqref{2d_4d_coupling_my}. Relation \eqref{2d_4d_coupling_my_noshift} between 4D and 2D couplings has been  obtained in \cite{Karasik} using localization. However, this formula uses an unconventional definition of the origin of the bulk $\theta_{4d}$ angle ($\theta_{4d}$ is shifted by $\pi$).
In this paper we  use the relation \eqref{2d_4d_coupling_my}.

\subsection{Dualities}
\label{dualities}

\begin{figure}[h]
	\centering
	\includegraphics[width=0.5\textwidth]{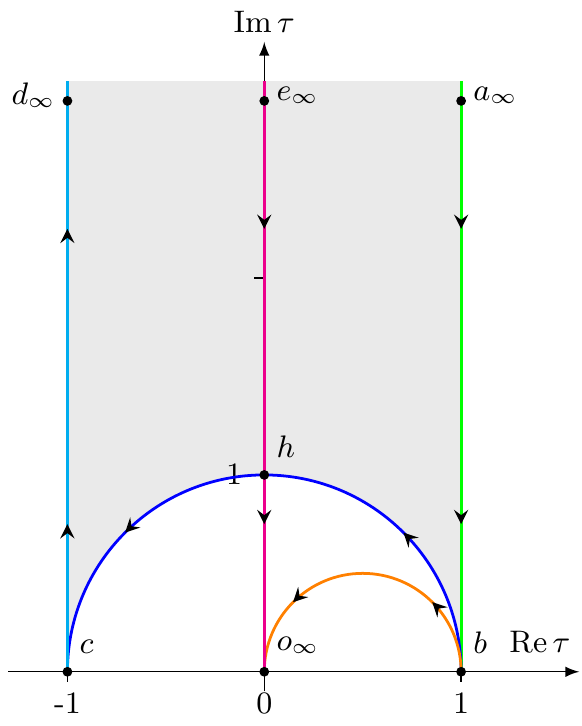}
	\caption{
		Fundamental domain of the duality group on the $\tau$ plane (shaded region).
		Shown are some particular trajectories in the space of the $\tau$ coupling. For the corresponding paths in the space of $\beta$ see Fig.~\ref{fig:beta_traj}. 
		The path $b \to o_\infty$ is an $ST^{-1}$ image of $b \to a_\infty$.
		The path $h \to o_\infty$ is an $S$-image of $h \to e_\infty$.
	}
\label{fig:coupling_traj}
\end{figure}

\begin{figure}[h]
	\centering
	\includegraphics[width=0.6\textwidth]{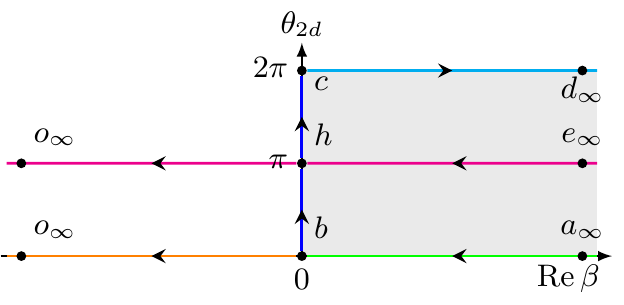}
	\caption{
		Fundamental domain of the duality group on the $\beta$ plane (shaded region).
		Shown are some particular trajectories in the space of the $\beta$ coupling. For the corresponding paths in the space of $\tau$ see Fig.~\ref{fig:coupling_traj}. 
		The path $b \to o_\infty$ is an $ST^{-1}$ image of $b \to a_\infty$.
		The path $h \to o_\infty$ is an $S$ image of $h \to e_\infty$.
		The trajectories are drawn modulo the relation $\theta_{2d} \sim \theta_{2d} + 2 \pi$.
	}
\label{fig:beta_traj}
\end{figure}

We have already seen that the world-sheet \wcpt model \eqref{wcp22} respects the duality transformation \eqref{2d_S-duality}. 
In this subsection we will see how this transformation is connected to the $S$ duality of the 4D SQCD, and discuss  other dualities as well.
We will define the $S$ duality and $T^\frac{1}{2}$ transformations as
\begin{equation}
\begin{aligned}
		S: \ \tau_\text{SW} &\to \frac{-1}{\tau_\text{SW}} \,,\\[2mm]
		T^\frac{1}{2}: \ \tau_\text{SW} &\to \tau_\text{SW} + 1 \,.
\end{aligned}
\label{ST_transformation}
\end{equation}
The conventional $T$ duality transformation $\tau_\text{SW} \to \tau_\text{SW} + 2$ is just a square of $T^\frac{1}{2}$.

The above transformations in Eq. \eqref{ST_transformation} generate the modular group SL($2,{\mathbb Z}$).
For the theory with the SU$(2)$ gauge group the  duality group is not a full SL($2,{\mathbb Z}$), but rather a subgroup generated by $S$ and $T$, the so-called $\Gamma^0(2)$ congruence subgroup of SL($2,{\mathbb Z}$).
It is not difficult to find the fundamental domain, see Fig.~\ref{fig:coupling_traj}.

In \cite{Karasik} it was shown  that 4D \ntwo SQCD with the U$(N)$ gauge group is not invariant under $S$ duality.
Say, our theory with the equal U(1) charges of four  quarks is mapped onto a SQCD with different U(1) quark charges. However, 
SQCD with the U(2) gauge group and equal U(1) charges  is invariant under the $ST^\frac{1}{2}S$ transformation \cite{Karasik}.  This transformation   in the world-sheet theory language  means that the theory is invariant under the sign change $\beta \to - \beta$. 

In our convention for the $\theta_{4d}$ angle, see Eq. \eqref{2d_4d_coupling_my}, the corresponding duality transformation is in fact 
an $S$ transformation.
Indeed, the $\theta_{4d}$ angle of \cite{Karasik} differs from ours by a shift by $\pi$ (cf. \textquote{+1} in \eqref{2d_4d_coupling_my}), which is a $T^\frac{1}{2}$ transformation). Since without this shift, the duality transformation would be $ST^\frac{1}{2}S$, our duality transformation is in fact
\begin{equation}
	T^\frac{1}{2} \cdot ST^\frac{1}{2}S \cdot T^\frac{1}{2} = S  \,.
\end{equation}
This identity can be checked explicitly.

Let us have a closer look at the $S$ duality.
Under the $S$ transformation the 4D coupling is transformed as
\begin{equation}
	\tau_\text{SW} \xrightarrow{S} \frac{-1}{\tau_\text{SW}} \,,
\end{equation}
and the $\lambda$ function in Eq. \eqref{2d_4d_coupling_my} becomes (see Eq. \eqref{lambda_p1_S}):
\begin{equation}
	\lambda(\tau_\text{SW} + 1) \xrightarrow{S}  \frac{1}{\lambda(\tau_\text{SW} + 1)} \,,
\end{equation} 
so that under the $S$ duality (cf. \eqref{2d_S-duality})
\begin{equation}
	\beta \xrightarrow{S} - \beta \,.
\label{selfdual_beta_eq}	
\end{equation}
Thus we have shown that the world-sheet duality \eqref{2d_S-duality} exactly corresponds to the $S$ duality of the bulk theory.
For an illustration, see Fig.~\ref{fig:coupling_traj} and~\ref{fig:beta_traj}.

The \wcpt self-dual point $\beta=0$ corresponds to $\tau_\text{SW} = 1$. Under the four-dimensional $S$-duality transformation, this maps to $\tau_\text{SW} = -1$, which differs from the initial value $\tau_\text{SW} = 1$ by a $2\pi$ shift of the $\theta_{4d}$ angle. The four-dimensional self-dual point $\tau=i$ corresponds to $\beta = i/2$ in two dimensions, see also Appendix~\ref{sec:dual_couplings}

%
%

\section{Conclusions \label{sec:conclusions}}

It has been known for a while now that the non-Abelian vortex string in four-dimensional \ntwo SQCD can become critical \cite{SYcstring}. 
This happens because, in addition to four translational zero modes of a usual ANO vortex, this string exhibits six orientational and  size zero modes. 
The target space of the effective world-sheet theory becomes $\mathbb{R}^4\times Y_6$, where $Y_6$ is a non-compact six-dimensional Calabi-Yau manifold, the so-called resolved conifold.

This has opened a way to quantize the solitonic string and to study the underlying gauge theory in terms of an \textquote{effective} string theory -- a kind of a \textquote{reverse holography} picture. 
It made possible quantitative description of the hadron spectrum \cite{KSYconifold,KSYcstring,SYlittles,SYlittmult}.
In particular, in \cite{SYlittles,SYlittmult}, the \textquote{Little String Theory} approach was used, namely a duality between the critical string  on the conifold 
and the non-critical $c=1$ string with the Liouville field and a compact scalar at the self-dual radius.
At the self-dual point $\beta=0$ of the world-sheet theory, the presence of the  massless 4D baryonic hypermultiplet $b$ was confirmed   and low-lying massive string states were also found.

In view of these spectacular results, the question arises: can we see these states directly from the field theory? 
In the present paper we managed to do just that.
To this end we employ the so-called 2D-4D correspondence. In the present case it means coincidence of the BPS spectra in the two-dimensional weighted sigma model, \wcpt \eqref{wcp22}, with the BPS spectrum in four-dimensional \ntwo SQCD with the U(2) gauge group  and four quark flavors  in the quarks vacuum. 
This coincidence was observed in \cite{Dorey,DoHoTo} and later explained 
in \cite{SYmon,HT2} using the picture of confined bulk monopoles which are seen as kinks in the 
world-sheet theory.
Then, we can reduce the problem to study of the BPS spectrum of the two-dimensional model {\em per se}.

Starting from weak coupling, we progressed into the strong coupling domain and further into the dual weak coupling domain. We managed to build a consistent picture of the BPS spectra in these regions and curves of marginal stability separating these domains. 
 
Consideration of the world-sheet kinks near the self-dual point $\beta=0$ led us to a rediscovery of a non-perturbative Higgs branch emerging at that point. The multiplet that lives on this branch turns out to be exactly the baryon multiplet $b$ found from 
string theory. 
Thus we have confirmed the consistency of the string theory picture describing the underlying gauge theory.

Moreover, in this model it was possible to observe the \textquote{instead-of-confinement} mechanism in action (see \cite{Shifman:2009mb,Shifman:2012yi} and a review \cite{SYdualrev}). At weak coupling $\beta \gg 1$ ($\beta$ being the sigma model coupling) there are perturbative states which look like $\mathbb{CP}(1)$ model excitations. At strong coupling $\beta \sim 1$ they decay into kink-antikink pairs. As we move further, we enter the dual weak coupling domain $\beta \ll -1$, with its own kinks and perturbative excitations. 
This evolution was described in the course of the present paper.

This world-sheet picture directly translates to the bulk theory. 
At weak coupling $g^2 \ll 1$ the perturbative spectrum of the four-dimensional \ntwo SQCD contains screened quarks and Higgsed gauge bosons. There are also solitonic states -- monopoles connected with non-Abelian flux tubes, forming mesons; but they are very heavy. As we progress into the strong coupling domain $g^2 \sim 1$, the screened quarks and Higgsed gauge bosons decay into confined monopole-antimonopole pairs. The \textquote{instead-of-confinement} phase is an alternative to the conventional confinement phase in QCD.

Similar instead-of-confinement phase appears if we move from large negative $\beta$ towards the strong coupling at $\beta \sim -1$.
In 4D SQCD this corresponds to moving  from the origin in the $\tau$-plane towards the upper semicircle shown in 
Fig.~\ref{fig:coupling_traj}. 
It is important  that $S$ dualities in the world-sheet and bulk theories are directly related, see Sec.~\ref{sec:2D_4D}.

\section*{Acknowledgments}

Useful discussions with E. Gerchkovitz and A. Karasik are acknowledged.
The work of MS is supported in part by DOE grant DE-SC0011842.
The work of A.Y. was  supported by William I. Fine Theoretical Physics Institute,   
University of Minnesota and 
by Russian Foundation for Basic Research Grant No. 18-02-00048a. 
The work of E.I. was supported in part by the Foundation for the Advancement of Theoretical Physics and Mathematics \textquote{BASIS} according to the research project No. \mbox{19-1-5-106-1}, and by Russian Foundation for Basic Research Grant No. 18-02-00048a.

%
%

\clearpage
\appendix

\begin{appendices}

\section{Secondary curves}
\label{sec:secondary_CMS}

Now we will investigate the decay curves of other particles and draw the corresponding CMS. Do this end, one has to keep in mind that the BPS kink central charge \eqref{BPSmass} is, generally speaking, a multi-branched function. 
On the $\beta$ plane, it can have branch cuts originating at the points where the kink mass develops some kind of a singularity. 
(This could be ignored while considering the primary CMS \eqref{CMS_right_curve} and \eqref{CMS_left_curve}, but not for the present task.)
From the explicit expressions for the kink mass we see that it is singular at the origin (see Eq. \eqref{kink_mass_P_beta0} and \eqref{kink_mass_K_beta0}) 
and at the AD points (see Eq. \eqref{CP1_Z_AD_2}). Therefore there are cuts originating from these points (modulo the $2\pi$ periodicity in the $\theta_{2d}$ direction).

\subsection{\textquote{Extra} kink decays}

When we go from strong coupling into the weak coupling domain $\beta \gg 0$, the kinks $[Z_P],\ P=1,2$ do not decay (they become massless at the AD points on the right curve \eqref{CMS_right_curve}, and they can be \textquote{dragged} through these points, where these kinks are the only massless particles and therefore absolutely stable \cite{Ferrari:1996sv}). On the other hand, the kinks  $[Z_K],\ K=3,4$ have masses of the order $~ |m_K - \ov{m}|$, and therefore they could decay into, say, a pair $[Z_P]$ + bifundamental. They can decay via the process
\begin{equation}
	[Z_K] \to [Z_P] + [- i (m_P - m_K)] \,.
\label{MK-MP_reaction}	
\end{equation}
In the dual weak coupling domain $\beta \ll 0$, the $P$-kinks decay via
\begin{equation}
	[Z_P] \to [Z_K] + [ i (m_P - m_K)] \,.
\label{MP-MK_reaction}	
\end{equation}

\begin{figure}[h]
	\centering
	\includegraphics[width=0.7\linewidth]{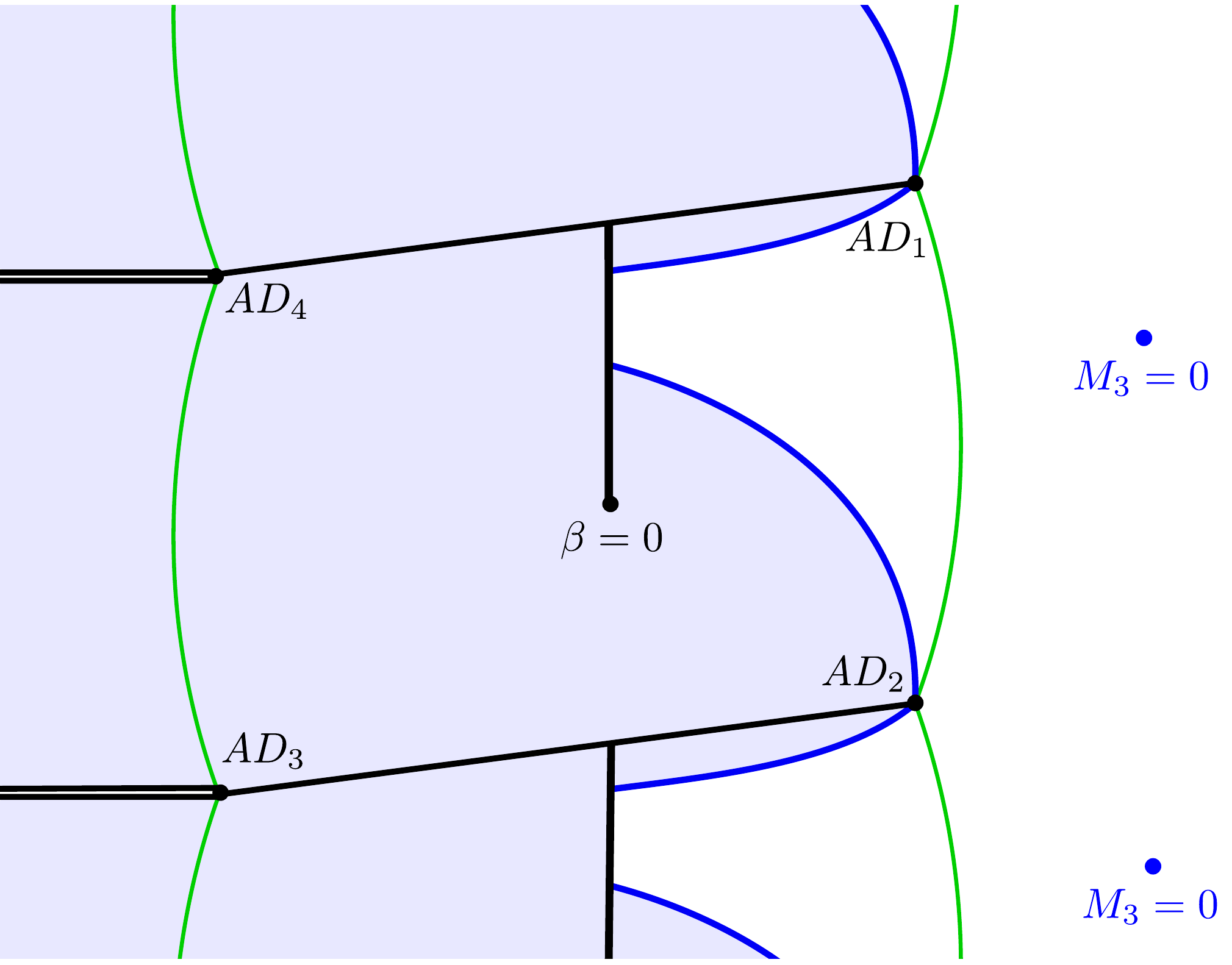}
	\caption{
		Complex plane of $\beta$.
		Schematic representation of the CMS structure for the $M_3$ kink decay. (Picture for $M_4$ is qualitatively the same.)
		Thin green lines are the primary curves.
		Thick black lines are the cuts.
		Thick blue lines are the CMS for the decays of $M_3$.
		Blue-shaded region is the domain of existence of $M_3$.
		The blue points on the right are where the mass of would-be $M_3$ state would vanish, see \eqref{M_K_large_beta}.
		}
\label{fig:CMS_right_13}
\end{figure}

Moreover, under some conditions these kinks must decay, otherwise they could become massless at some points at weak coupling. 
To see this, consider a $K$-kink central charge at weak coupling in the CP(1) limit \eqref{CP1_limit}. Formula \eqref{CP1_Z_quasiclassical} can be straightforwardly generalized for $K$-kinks as
\begin{equation}
	Z_K \approx  - \beta_{CP(1)} \cdot \delta m_{12} + i \, (m_K - \ov{m})  + \frac{\delta m_{12}}{\pi} \,,
\end{equation}
so that the mass of the corresponding state is the limit $\Re\beta \equiv r \gg 1$ is given by
\begin{equation}
	M_K \approx |\delta m_{12}| \cdot \left( r - \frac{1}{\pi} \ln \abs{\frac{\Delta m }{\delta m_{12}}} - \frac{1}{\pi} - \Im\frac{m_K - \ov{m}}{\delta m_{12}} \right) \,.
\label{M_K_large_beta}
\end{equation}

We see that for certain mass choices there are points at weak coupling (r.h.s. domain $\beta \gg \beta_{AD}$) where $M_K$ vanish . Therefore, these states must decay.
Analogously, $M_P$  kinks ($P=1,2$) may become massless in the dual weak coupling domain $\beta \ll - \beta_{AD}$. Their mass in this limit is given by
\begin{equation}
	M_P \approx |\delta m_{34}| \cdot \left( r - \frac{1}{\pi} \ln \abs{\frac{\Delta m }{\delta m_{34}}} - \frac{1}{\pi} + \Im\frac{m_P - \wt{m}}{\delta m_{34}} \right) \,.
\end{equation}

The  CMS equation for both decays \eqref{MK-MP_reaction} and \eqref{MP-MK_reaction} is
\begin{equation}
	\Re \left( \frac{ Z_P }{m_P - m_K} \right) = 0  
	\Leftrightarrow
	\Re \left( \frac{ Z_K }{m_P - m_K} \right) = 0 \,.
\label{CMS_K_decay}
\end{equation}
We must add to this equation  the condition that a particle cannot decay into heaver particles:
\begin{equation}
\begin{aligned}
	| Z_K | &= | Z_P | + | - i ( m_P - m_K ) | \quad \text{for the decay \eqref{MK-MP_reaction}} \,, \\[2mm]
	| Z_P | &= | Z_K | + |   i ( m_P - m_K ) | \quad \text{for the decay \eqref{MP-MK_reaction}} \,.
\end{aligned}
\label{abs_decay_condition}
\end{equation}
In the case when $m_1,\ m_2,\ m_K$ for some $K$ lie on a straight line in the complex plane, the CMS for the decay \eqref{MK-MP_reaction} {\em coincides with the primary curve} \eqref{CMS_right_curve}. Conversely, when for some $P$ the masses $m_P,\ m_3,\ m_4$ are aligned, the CMS for \eqref{MP-MK_reaction} coincides with the dual primary curve \eqref{CMS_left_curve}.
%

The CMS equation \eqref{CMS_K_decay} simplifies in the CP(1) limit \eqref{CP1_limit}. 
Using a simple generalization of the approximate central charge formula \eqref{CP1_Z_AD_2} we can rewrite this equation near an AD point as:
\begin{equation}
	\Re \left[ \frac{m_1 - m_2}{m_P - m_K} \cdot (\beta - \beta_{AD_P})^{3/2} \right] = 0
\end{equation}
for  $P=1,2$. (The indices of AD points follow Fig.~\ref{fig:CMS_type1}.) This equation is equivalent to
\begin{equation}
	\cos \left( \frac{3}{2} \arg(\beta - \beta_{AD_P}) + \phi_{PK} \right) = 0
	\,, \quad
	\phi_{PK} = \arg\left( \frac{m_1 - m_2}{m_P - m_K} \right) \,.
\end{equation}
The solution is represented by lines originating from the $AD_P$ point and going out at angles
\begin{equation}
	\arg(\beta - \beta_{AD_P}) = -  \frac{2}{3} \phi_{PK} - \frac{\pi}{3} + \frac{2}{3} \pi \, n
	\,, \quad
	n \in \mathbb{Z} \,.
\label{CMS_13_near_AD1}	
\end{equation}
From this equation we see that, generally speaking, three different CMS originate in the Argyres-Douglas point $AD_P$ (this is very similar to the CP(1) case). However, only some of them satisfy the additional condition \eqref{abs_decay_condition}. Namely, for \eqref{abs_decay_condition} to be true, we have to impose
\begin{equation}
	\arg{Z_P} - \arg(-i(m_P - m_K))  \in 2 \pi \mathbb{Z} \,.
\end{equation}
This condition leaves us with even $n$ in \eqref{CMS_13_near_AD1}.
These are the CMS for the decay \eqref{MK-MP_reaction} near the point $AD_P$. 

Let us look at the equation \eqref{CMS_13_near_AD1} more closely. 
Depending on $\phi_{PK}$, the qualitative picture changes. When it is zero, then the CMS is stretched between $AD_1$ and $AD_2$ and coincides with the primary curve \eqref{CMS_right_curve}.
When $\phi_{PK} \in [-\pi/2, 0)$, the CMS bends into the strong coupling domain, and the kink $M_K$ cannot penetrate into the weak coupling domain at $\beta > 0$.
If $\phi_{PK} \in (0, \pi/2]$, the CMS for the $K$-kink decay goes into the weak coupling domain, and the $K$-kink is present in some subregion of the weak coupling at $\beta > 0$. But of course it cannot reach the region where its mass would vanish, see \eqref{M_K_large_beta}. Other values of $\phi_{PK}$ are recovered by relabeling $1 \leftrightarrow 2$. On Fig.~\ref{fig:CMS_right_13} we present the CMS for the $[Z_3]$ kink decays (CMS for decays of $[Z_4]$ is qualitatively the same).
 
The same argument can be applied to the $P$-kinks near the dual weak coupling domain at $\beta < 0$. If $\arg\left( (m_4 - m_3) / (m_P - m_K) \right)$ is zero, the CMS for the decay \eqref{MP-MK_reaction} coincides with the dual primary curve \eqref{CMS_left_curve}. When this $\arg$ is positive, the $P$-kinks cannot penetrate into the dual weak coupling region. When it is negative, the $P$-kinks are present in a subregion of the dual weak coupling domain, but they never reach the regions where their masses would vanish.

\subsection{Decay of strong coupling tower of  higher winding states 
\label{sec:higher_winding}}

Now we briefly discuss decays of the $n \neq 0$ states of the tower \eqref{Z_higher_windings}.
In the limit $\Delta m \gg \delta m_{12} \,, \ \delta m_{34}$ they can decay into the states of lower winding with emission of bifundamentals. For example, if $n > 0$, some of the decays are
\begin{equation}
\begin{aligned}
	[Z_1^{[n]}] &~\to~ [Z_4^{[ n ]}] ~+~ [i (m_1 - m_4)]  \,, \\[2mm]
	[Z_4^{[n]}] &~\to~ [Z_2^{[ n-1 ]}] ~+~ [i (m_1 - m_3)]  \,.  
\end{aligned}
\end{equation}
More generally, the $n > 0$ states can decay as
\begin{equation}
\begin{aligned}
	[Z_P^{[n]}] &~\to~ [Z_K^{[ n ]}] ~+~ [i (m_P - m_K)]  \,, \\[2mm]
	[Z_K^{[n]}] &~\to~ [Z_P^{[ n-1 ]}] ~+~ [i (m_{\wt{P}} - m_{\wt{K}})]  \,,
\end{aligned}
\end{equation}
where $P, \wt{P}$ is some permutation of indices $1,\ 2$, and $K, \wt{K}$ is a permutation of $3,\ 4$. The states with $n < 0$ decay similarly.

The corresponding CMS satisfies the equation
\begin{equation}
	\Re \left( \frac{ Z_P^{[n]} }{m_P - m_K} \right) = 0  
	\,, \quad 
	\Re \left( \frac{ Z_K^{[n]} }{m_{\wt{P}} - m_{\wt{K}}} \right) = 0 \,.
\label{CMS_higherwind_decay}
\end{equation}
Far from the origin, when $\beta \gg 1$ in the $\mathbb{CP}(1)\;$ limit \eqref{CP1_limit}, the equations \eqref{CMS_higherwind_decay} differ from \eqref{CMS_K_decay} only by $O(\delta m_{12} / \Delta m ,\, \delta m_{34} / \Delta m)$ terms; therefore, the corresponding CMS should be close on each other, at least in some region.

Careful numerical studies show that there are two possibilities: either the CMS \eqref{CMS_higherwind_decay} form closed curves lying inside the strong coupling domain, or they form spirals that go to the origin. In any case, it follows that the higher winding states considered here live exclusively inside the strong coupling domain and cannot get into the weak coupling regions.

\section{Modular functions}
\label{sec:modular_101}

\subsection{$\theta$ functions}

Let us introduce the nome
\begin{equation}
	q = e^{i \pi \tau_\text{SW}} = e^{2 i \pi \tau}
\label{nome}	
\end{equation}
where $\tau_\text{SW}$ is the gauge coupling defined in \eqref{tau_def}.
We define $\theta$-functions as in \cite{SW2}. In terms of the nome \eqref{nome} they are
\begin{equation}
\begin{aligned}
	\theta_1 (q) &= \sum_{n \in \mathbb{Z}} q^{(n + 1/2)^2}		= 2 q^{1/4} (1 + q^2 + \ldots)  \,, \\
	\theta_2 (q) &= \sum_{n \in \mathbb{Z}} (-1)^n \, q^{n^2}	= 1 - 2 q + \ldots  \,, \\
	\theta_3 (q) &= \sum_{n \in \mathbb{Z}} q^{n^2}				= 1 + 2 q + \ldots  \,. \\
\end{aligned}
\label{theta_def}
\end{equation}
There are many relations among these, e.g. \cite{Chandrasekharan}
\begin{equation}
	\theta_3^4 = \theta_2^4 + \theta_1^4 \,.
\end{equation}

The $\theta$ functions \eqref{theta_def} are obviously invariant under $T$ transformation $\tau_\text{SW} \to \tau_\text{SW} + 2$ \eqref{ST_transformation}. Moreover, the following identities \cite[eq. (8.10)]{Chandrasekharan} hold for the $T^\frac{1}{2}$ \eqref{ST_transformation} transformation:
\begin{equation}
\begin{aligned}
	\theta_1 \left( \tau_\text{SW} + 1 \right) &= e^\frac{i \pi}{4} \ \theta_1(\tau_\text{SW})  \,,  \\[2mm]
	\theta_2 \left( \tau_\text{SW} + 1 \right) &= \theta_3(\tau_\text{SW})  \,, \\[2mm]
	\theta_3 \left( \tau_\text{SW} + 1 \right) &= \theta_2(\tau_\text{SW})  \,. \\
\end{aligned}
\label{half_shift}
\end{equation} 
Under $S$ \eqref{ST_transformation}, we have \cite[eq. (8.9)]{Chandrasekharan}
\begin{equation}
\begin{aligned}\\
	\theta_1 \left(- \frac{1}{\tau_\text{SW}} \right) &= \sqrt{- i \tau_\text{SW}} \ \theta_2(\tau_\text{SW})  \,, \\[2mm]
	\theta_2 \left(- \frac{1}{\tau_\text{SW}} \right) &= \sqrt{- i \tau_\text{SW}} \ \theta_1(\tau_\text{SW})  \,, \\[2mm]
	\theta_3 \left(- \frac{1}{\tau_\text{SW}} \right) &= \sqrt{- i \tau_\text{SW}} \ \theta_3(\tau_\text{SW})  \,, \\
\end{aligned}
\label{theta_S_APS}
\end{equation} 
where $\sqrt{-i \tau_\text{SW}} = +1$ for $\tau_\text{SW} = i$.
Here we slightly abused notation by using the same letter $\theta$ as in \eqref{theta_def}.

\subsection{The \boldmath{$h$} function}

From the $\theta$ functions we can build modular functions. In the SW curve \eqref{4d_curve_sun} the $h$ function was used, which is defined as \cite{APS}
\begin{equation}
	h(\tau_\text{SW}) = \frac{2 \theta_1^4(\tau_\text{SW})}{\theta_2^4(\tau_\text{SW}) - \theta_1^4(\tau_\text{SW})}
\label{h_def_APS}	
\end{equation}
or, in terms of the nome \eqref{nome},
\begin{equation}
	h(q) = 32 \, q + O(q^2) \,.
\end{equation}
The $S$-transformation acts on \eqref{h_def_APS} as
\begin{equation}
	h\left(- \frac{1}{\tau_\text{SW}} \right) = - 2 - h(\tau_\text{SW}) \,,
\end{equation}
and the combination
\begin{equation}
	h \cdot (h + 2) = \frac{4 \theta_1^4 \theta_2^4}{(\theta_2^4 - \theta_1^4)^2}
\label{h_combination}	
\end{equation}
is invariant with respect to $S$ and $T$ transformations. Under the half shift $T^\frac{1}{2}$ it becomes
\begin{equation}
	h(\tau_\text{SW} + 1) \cdot (h(\tau_\text{SW} + 1) + 2)  
		= - \frac{4 \theta_3^4(\tau_\text{SW}) \cdot \theta_2^4(\tau_\text{SW})}{(\theta_3^4(\tau_\text{SW}) + \theta_1^4(\tau_\text{SW}))^2}
		\,.
\end{equation}

\subsection{The \boldmath{$\lambda$} function}
\label{klambdaf}

In this paper we have also used the modular $\lambda$ function (see e.g. \eqref{2d_4d_coupling_my}) which can be expressed as 
\begin{equation}
	\lambda(\tau_\text{SW}) 
		= \frac{\theta_1^4(\tau_\text{SW}) }{\theta_3^4(\tau_\text{SW}) }
		= 16 q - 128 q^2 + O(q^3)
\label{lambda_function}		
\end{equation}
where $q$ is the nome \eqref{nome}.
This function is again invariant under the $T$ transformation, while under $S$ it transforms as
\begin{equation}
	\lambda\left(- \frac{1}{\tau_\text{SW}} \right) = 1 - \lambda(\tau_\text{SW}) \,.
\label{lambda_trans_S}	
\end{equation}
Under the half shift \eqref{half_shift} this becomes
\begin{equation}
	\lambda(\tau_\text{SW} + 1) = \frac{\lambda(\tau_\text{SW})}{\lambda(\tau_\text{SW}) - 1} = - \frac{\theta_1^4(\tau_\text{SW}) }{\theta_2^4(\tau_\text{SW}) } \,.
\label{lambda_trans_halfshift}	
\end{equation} 
From \eqref{lambda_trans_S} and \eqref{lambda_trans_halfshift} we see that under the $S$ transformation
\begin{equation}
	\lambda(\tau_\text{SW} + 1) \xrightarrow{S} \lambda\left(- \frac{1}{\tau_\text{SW}} + 1 \right) = \frac{1}{\lambda(\tau_\text{SW} + 1)} \,.
\label{lambda_p1_S}	
\end{equation} 
Using \eqref{h_combination} and \eqref{lambda_trans_halfshift}	we can write down a relation between $\lambda$ and $h$ function,
\begin{equation}
	- h(\tau_\text{SW})[h(\tau_\text{SW}) +2] = \frac{4 \, \lambda(\tau_\text{SW} + 1)}{(1 + \lambda(\tau_\text{SW} + 1))^2} \,.
\label{h_tau_relation}	
\end{equation}

The inverse of $\lambda(\tau)$ is given in terms of the hypergeometric functions
\begin{equation}
	  \tau = i~\frac{{}_2F_1\left(1/2,1/2;1;1-\lambda\right)}{ {}_2F_1\left(1/2,1/2;1;\lambda\right)} \,.
\label{lambda_inv_2F1}	  
\end{equation}
In terms of the complete elliptic integral of the first kind $K(k)$, 
\begin{equation}
	  \tau = i~\frac{K(\sqrt{1 - \lambda})}{K(\sqrt{\lambda})} \,.
\label{lambda_inv_K}	  
\end{equation}

%
%

\section{Central charge windings at strong coupling \label{sec:Z_windings}}

In this section we are going to derive different windings ot the central charge \eqref{BPSmass} indicated on Fig.~\ref{fig:CMS_type1}.

\begin{figure}[h]
    \centering
    \begin{subfigure}[t]{0.5\textwidth}
        \centering
        \includegraphics[width=\textwidth]{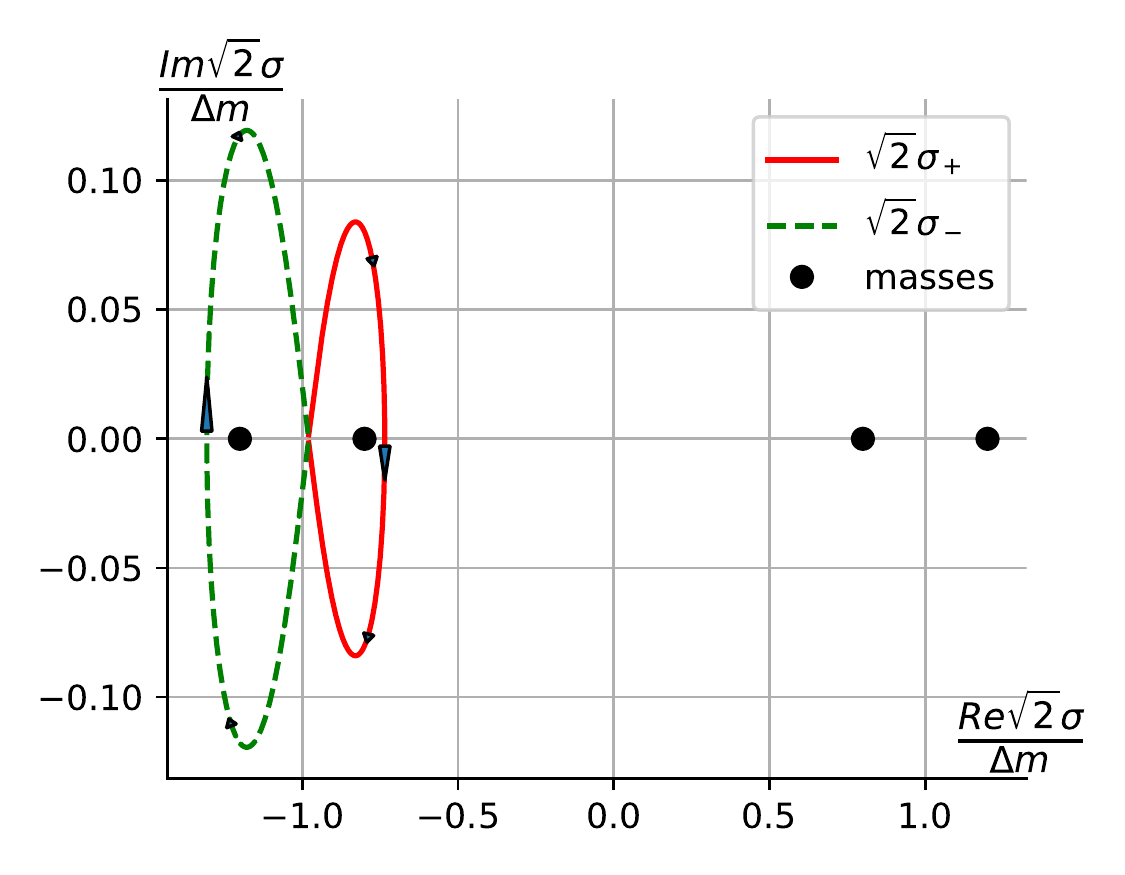}
        \caption{Trajectory of 2D roots $\sigma$ along \eqref{beta_traj_along_theta}}
        \label{fig:Z_windings_Imbeta}
    \end{subfigure}%
    ~
    \begin{subfigure}[t]{0.5\textwidth}
        \centering
        \includegraphics[width=\textwidth]{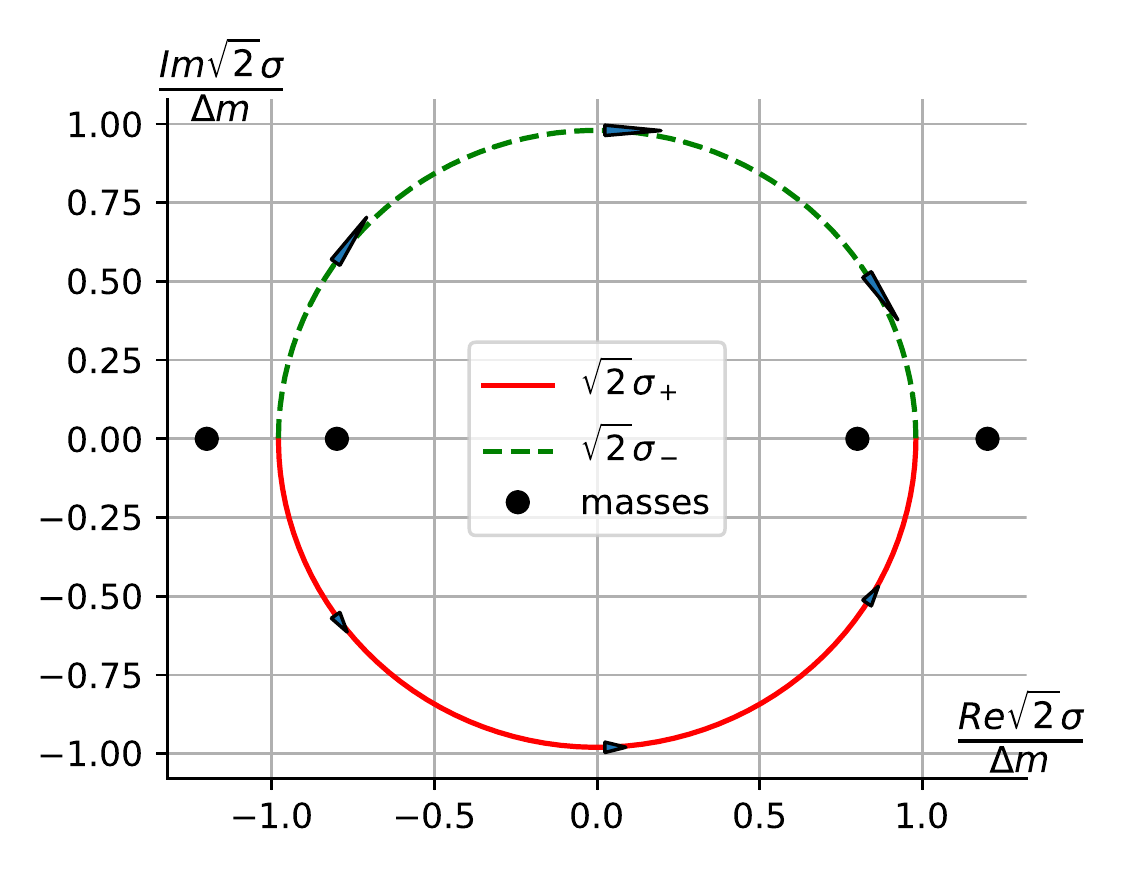}
        \caption{Trajectory of 2D roots $\sigma$ along \eqref{beta_traj_alongRe}}
        \label{fig:Z_windings_Rebeta}
    \end{subfigure}%
\caption{
	Trajectories of 2D roots $\sigma$ along different $\beta$-paths.
	Numerical results.
	Complex plane of $\sqrt{2}\sigma$.
	We see that $\sigma$-roots encircle the masses $m_A$ (represented by bullets).
}
\label{fig:Z_windings}
\end{figure}

\subsection{Winding along \boldmath{$\theta_{2d}$} \label{sec:along_theta}}

Now we are going to derive the $AD_1 \to AD_2$ phase shift from Fig.~\ref{fig:CMS_type1}. 
For simplicity we consider the CP(1) limit \eqref{CP1_limit}.
Positions of AD points $AD_1$ and $AD_2$ are approximately given by \eqref{AD_roots_simplecase_beta1_approx}.
Consider a trajectory in the $\beta$ plane, where the coupling flows continuously from one AD point at $\beta_{AD}$ to another at $\beta_{AD} + i$
\begin{equation}
	\beta = \frac{1}{\pi} \ln\frac{2 \, \Delta m}{\delta m_{12}}   
		+ \frac{i (t - \pi)}{2 \pi} -\varepsilon
		\, ,
		\quad
		1 \gg \varepsilon > 0 \, , \  t \in [0, 2\pi] \,.
\label{beta_traj_along_theta}		
\end{equation}
Here, $\varepsilon$ is just a regularization parameter.
Then, 
\begin{equation}
	e^{-2 \pi \beta} \approx - \left( \frac{\delta m_{12}}{2 \, \Delta m} \right)^2 \, e^{- i t} \, (1 + 2 \pi \varepsilon) \,,
\end{equation}
and for the expression under the square root in \eqref{roots_symmetric} (i.e. the discriminant) we get
\begin{equation}
	D \approx \frac{\delta m_{12}^2}{4} + \Lambda_{CP(1)}^2 =\frac{\delta m_{12}^2}{4} \, \left( 1 - (1 + \wt{\varepsilon}) e^{- i t} \right)
	\,, \quad
	1 \gg \wt{\varepsilon} > 0 \,.
\end{equation}
This expression winds around $1$ with the radius $(1 + \wt{\varepsilon})$ clockwise. Then the $\sigma$ vacua, which are approximately given by
\begin{equation}
	\sqrt{2}\sigma_\pm \approx  \pm \frac{\Delta m}{2} + \sqrt{D} \,,
\label{temp_xpm}
\end{equation}
wind, see Fig.~\ref{fig:Z_windings_Imbeta}. In the limit $\varepsilon \to 0$, the root $\sqrt{2}\sigma_+$ winds around $(- \Delta m + \delta m_{12})/2 = - m_2$ clockwise, while $\sqrt{2}\sigma_-$ winds around $(- \Delta m - \delta_{12}) / 2 = - m_1$ clockwise, both with radius $\delta m_{12} \, (1 + \wt{\varepsilon} / 2)$.

This yields nontrivial phase shifts in the mirror variables, see \eqref{mirror-x_map}. While $Y$'s stay intact, the $X_1$ winds because of $\sqrt{2}\sigma_-$ and picks up $- 2 \pi i$. $X_2$ winds because of $\sqrt{2}\sigma_+$ and picks up $- 2 \pi i$. Then, the complexified kink central charge defined by $Z = 2 ({\cal W}_{\rm mirror}(Vac_-) - {\cal W}_{\rm mirror}(Vac_+))$ is shifted by $- i (m_1 - m_2)$. Therefore if $Z_2=0$  at $AD_2$ than $Z_1=Z_2 +i(m_1-m_2)$ 
becomes zero    at  $AD_1$, see \eqref{kink_mass_P_CP1_1}. In other words $[Z_1]$ and $[Z_2]$ kinks are massless at AD points $AD_1$ and $AD_2$ respectively.

With the same reasoning we can prove the $AD_3 \to AD_4$ shift from Fig.~\ref{fig:CMS_type1}.

\subsection{From positive to negative \boldmath{$\beta$}}

Now, consider the trajectory in the $\beta$ plane going from right to left, i.e. from $AD_2$ to $AD_3$ on Fig.~\ref{fig:CMS_type1}.
For simplicity we consider the limit of real $\Delta m \gg \delta m_{12} = \delta m_{34} > 0$.
\begin{equation}
	\beta \approx t \left[ \frac{1}{\pi} \ln\frac{2 \,  \Delta m}{\delta m_{12}}   - \varepsilon \right] - \frac{i}{2}
		\, ,
		\quad
		1 \gg \varepsilon > 0 \, , \  t \in [1, -1]
		\,.
\label{beta_traj_alongRe}
\end{equation}
When $t$ changes from $1$ to $-1$, the value of $\beta$ flows from $AD_2$ to $AD_3$. Then we have
\begin{equation}
	e^{- 2 \pi \beta} \approx - \left(\frac{\delta m_{12}}{2 \, \Delta m}\right) ^ {2 t} \ (1 + 2 \pi t \varepsilon) \,,
\end{equation}
and for the expression under the square root in \eqref{roots_symmetric} (i.e. the discriminant) we get
\begin{equation}
	D \approx \delta m_{12}^2 \left( 
		1 - \frac
				{\left( \frac{2 \, \Delta m}{\delta m_{12}} \right) ^ {2 (1-t)} \ (1 + 2 \pi t \varepsilon)}
				{\left( 1 + \left( \frac{\delta m_{12}}{2 \, \Delta m} \right) ^ {2 t} \ (1 + 2 \pi t \varepsilon) \right)^2} 
	\right) \,.
\label{temp_discr}
\end{equation}
When $\delta m_{12} < \Delta m$, this expression always gives negative $D$. 
There is no nontrivial windings of the roots. However the first term in the root formula \eqref{roots_symmetric} changes smoothly from $- \Delta m / 2$ to $+ \Delta m / 2$ as $t$ is varied from $1$ to $-1$. Therefore, both SW roots evolve from the vicinity of $- \Delta m / 2$ at $\beta \sim AD_2$ to the vicinity of $+ \Delta m / 2$ at $\beta \sim AD_3$. Turns out that in terms of $\sqrt{2}\sigma_\pm$ from \eqref{temp_xpm}, the root $\sqrt{2}\sigma_+$ travels in the lower half plane, while the root $\sqrt{2}\sigma_-$ travels in the upper half plane, see Fig.~\ref{fig:Z_windings_Rebeta}.

From this and the map \eqref{mirror-x_map} it follows that, when $\beta$ flows from $AD_2$ to $AD_3$, the mirror variable $X_1$ stays in the right half plane $\Re X_1 > 0$, $Y_4$ stays in the left half plane $\Re Y_4 < 0$. $X_2$ and $Y_3$ each pick up $+i \pi$ because of the $\sigma_+$ change, while because of the $\sigma_-$ they each pick up $- i \pi$. All in all, the complexified kink central charge defined by $Z = 2 ({\cal W}_{\rm mirror}(Vac_-) - {\cal W}_{\rm mirror}(Vac_+))$ is shifted by $- i (m_2 - m_3)$, which is exactly the shift indicated on Fig.~\ref{fig:CMS_type1}. Thus $[Z_2]$ and $[Z_3]$ kinks are massless at AD points $AD_2$ and $AD_3$ respectively.

With the same reasoning we can prove the $AD_4 \to AD_1$ monodromy from Fig.~\ref{fig:CMS_type1}. Also, it is consistent with the $\mathbb{Z}_2$ transformation, see Fig.~\ref{fig:CMS_type1}.

%
%

\section{More on self-dual couplings \label{sec:dual_couplings}}

Consider 4d self-dual points. Corresponding $\tau_\text{SW}$ should satisfy the equation
\begin{equation}
	\tau_\text{SW} = \frac{-1}{\tau_\text{SW}} \,.
\label{selfdual_eq_naive}	
\end{equation}
The solution in the upper half plane is
\begin{equation}
	\tau_0 = i \,.
\label{selfdual_tau_naive}	
\end{equation}
However, if we take into account also $T$ duality, then the equation \eqref{selfdual_eq_naive} is modified:
\begin{equation}
	\tau_\text{SW} = \frac{-1}{\tau_\text{SW}} + 2 \, k \,,
	\quad
	k \in {\mathbb Z} \,.
\label{selfdual_eq_T}	
\end{equation}
Solving this equation, we obtain a whole series of self-dual points,
\begin{equation}
	\tau_{\pm k} = k \pm \sqrt{k^2 - 1}\,,
	\quad
	k \in {\mathbb Z} \,,
\end{equation} 
or, equivalently,
\begin{equation}
	\tau_{\pm k} = \pm (k - \sqrt{k^2 - 1} ) \,,
	\quad
	k \in \{0, \, 1, \, 2, \, \ldots \} \,.
\label{selfdual_tau_T}		
\end{equation} 
For $k=0$ this gives \eqref{selfdual_tau_naive}. For $k=1$, this gives a point 
\begin{equation}
	\tau_1 = 1 \,.
\label{selfdual_tau=1}
\end{equation}

Now consider the 2D self-dual points. An obvious point \eqref{selfdual_beta_eq} is
\begin{equation}
	\beta_0 = 0 \,, 
	\quad
	e^{-2 \pi \beta_0} = +1 \,.
\label{selfdual_beta_naive}
\end{equation}
But if we take into account the 2d $T$ duality $\beta \to \beta + i$, we see that in fact there is 
a whole series of the points self-dual under $S$ \eqref{selfdual_beta_eq},
\begin{equation}
	\beta_k =  \frac{i}{2} \, k\,, 
	\quad
	e^{-2 \pi \beta_1} = (-1)^k \,,
	\quad
	k \in \mathbb{Z} \,.
\label{selfdual_beta_T}	
\end{equation}
We have seen some of them in Sec.~\ref{sec:2D_4D}:
\begin{equation}
\begin{aligned}
	\tau_0 = i &\leftrightarrow \beta_1 =  \frac{i}{2}  \,,\\[2mm]
	\tau_1 = 1 &\leftrightarrow \beta_0 = 0  \,.\\
\end{aligned}	
\end{equation}

\end{appendices}

%
%


\clearpage


\end{document}